\documentclass{jfm}
\usepackage{graphicx} 

\usepackage{graphicx}
\usepackage{epstopdf, epsfig}
\usepackage{xcolor}
\usepackage{bm}
\usepackage{physics}
\usepackage{tabularx}
\usepackage{wasysym}
\usepackage{tikz}
\usepackage{hyperref}

\title{Granular segregation across flow geometries: a closure model for the particle segregation velocity}

\colorlet{blue}{black}
\colorlet{red}{black}
\colorlet{orange}{black}

\author{Yifei Duan\aff{1},
 Lu Jing\aff{2,3},
 Paul B. Umbanhowar\aff{4},
 Julio M. Ottino\aff{1,4,5},
\and Richard M. Lueptow\aff{1,4,5}
 \corresp{\email{r-lueptow@northwestern.edu}}
}
\affiliation{\aff{1}Department of Chemical and Biological Engineering, Northwestern University, Evanston, Illinois 60208, USA
\aff{2}Institute for Ocean Engineering, Shenzhen International Graduate School, Tsinghua University, Shenzhen 518055, China\
\aff{3}State Key Laboratory of Hydroscience and Engineering, Tsinghua University, Beijing, 100084, China\
\aff{4}Department of Mechanical Engineering, Northwestern University, Evanston, Illinois 60208, USA
\aff{5}Northwestern Institute on Complex Systems (NICO), Northwestern University, Evanston, Illinois 60208, USA
}

\begin{document}

\maketitle

\begin{abstract}
Predicting particle segregation has remained challenging due to the lack of a general model for the segregation velocity that is applicable across a range of granular flow geometries. Here, a segregation velocity model for dense granular flows is developed by exploiting force balance and recent advances in particle-scale modelling of the segregation driving and drag forces over \textcolor{blue}{the entire particle concentration range, size ratios up to 3, and inertial numbers as large as} \textcolor{red}{0.4.} This model is shown to correctly predict particle segregation velocity in a diverse set of idealized and natural granular flow geometries simulated using the discrete element method. When incorporated in the well-established advection-diffusion-segregation formulation, the model has the potential to accurately capture  segregation phenomena in many relevant industrial applications and geophysical settings.
\end{abstract}

\section{Introduction}

A classic problem in fluid mechanics is the terminal velocity of a spherical particle in an otherwise quiescent viscous liquid. A balance of weight, buoyancy, and Stokes drag, which depends on the velocity of the particle, yields the terminal velocity. While the result is affected by nearby particles, adjacent walls, etc., the essence of the problem resides in the force balance on the particle. 

\textcolor{blue}{In this paper, we consider the analogous problem in granular shear flow and use it to determine the segregation velocity in mixtures of two particle species having different sizes. To do this, we use a force balance of the particle weight, the segregation force, and the granular drag force to determine the terminal velocity, or, equivalently, the segregation velocity, of an intruder particle in a granular flow and then extend this to the segregation velocity in mixtures of two particle species. However, the construction of the equivalent ``granular terminal velocity" problem requires a crucial addition compared to the fluid problem.} An intruder particle in a granular system will only move if energy, e.g., in the form of vibration or shear, is added to the system. \textcolor{blue}{In the case of shear, which is considered here,} the local shear profile in the intruder vicinity affects the forces acting on it. Furthermore, the buoyancy force on a particle in a granular medium differs somewhat from that for a particle in a fluid. The forces related to shear and buoyancy can be expressed in terms of a ``segregation force" that includes both kinematics- and gravity-dependent terms~\citep{jing_unified_2021}.  Last, the relationship between the drag force and the particle velocity in a granular flow has recently been clarified to be Stokesian in character \citep{tripathi2013density,jing_drag_2022}. \textcolor{blue}{Thus,} a force balance of the particle weight, the segregation force, and the granular drag force \textcolor{blue}{can be used to determine the segregation velocity for an intruder particle over} a wide range of granular flow conditions.

More generally and more significantly, the granular terminal velocity of an intruder particle can be connected to the problem of particle size segregation in a flowing mixture of small and large particles with arbitrary finite concentrations. In the mixture, small particles fall through interstices between large particles to lower parts of a flowing layer, thereby forcing large particles upward and resulting in the spatial segregation of initially mixed small and large particles~\citep{gray2018particle,umbanhowar2019modeling}. If segregation of two particle species is viewed as one of the two species migrating relative to the other within the bulk flow\textcolor{orange}{~\citep{bridgwater1985particle, savage1988particle}}, the segregation velocity of each species \textcolor{orange}{can be thought of as} a granular terminal velocity problem. In this case, the segregation force and drag force models need to be extended from the single-intruder limit to mixtures of arbitrary species concentrations. Such extensions bridging particle-level forces and continuum models of the segregation velocity are the focus of recent work \citep{rousseau_bridging_2021,tripathi_theory_2021,sahu_particle_2023,duan_segregation_2022,duan2024general}, leading to the emergence of a new physics-based approach to segregation velocity modelling. The present paper aims to complete this new approach and demonstrate its general applicability using a wide range of granular flow configurations.  


Modelling particle size segregation in the continuum framework usually involves solving the spatial and temporal evolution of particle concentration via an advection-diffusion-segregation equation, \textcolor{blue}{first suggested by~\cite{bridgwater1985particle},} which differs from a standard advection-diffusion formulation in that a closure relation is needed for the segregation velocity~\citep{gray2018particle,umbanhowar2019modeling,thornton_brief_2021}. Therefore, how to model the segregation velocity is the central question of most recent segregation theories. \textcolor{blue}{Several different approaches have been used to understand, model, and predict the segregation velocity. In one approach, the segregation velocity} is directly determined from discrete element method (DEM) simulations~\citep{umbanhowar2019modeling}, which empirically connect the local segregation velocity with flow kinematics (shear rate), species concentration, and relative particle properties (size ratio and density ratio). Despite the semi-empirical nature of this approach, it has proved effective in capturing  segregation fluxes \citep{fan2014modelling,schlick2015modeling,jones2018asymmetric}, density- or shape-induced segregation \citep{xiao_modelling_2016,duan_modelling_2021,zhao2018simulation}, and fluid effects in segregation \citep{cui_particle_2022}, as well as segregation involving both gravity- and shear-gradient-related driving mechanisms \citep{fan_phase_2011,liu2023coupled,singh2024continuum}. The drawback of such empirical segregation velocity models is the lack of a universal model suitable for a wide range of particle properties and flow configurations.

\textcolor{blue}{The segregation velocity can also be extracted from particle-species-specific momentum equations (force balance at the continuum level) in a flowing mixture \citep{gray2018particle,thornton_brief_2021}.} In this framework, the \textcolor{red}{intrinsic} pressure gradient of a particle species drives segregation and is counteracted by inter-species drag and diffusive remixing \citep{gray2005theory,gray2006particle,bancroft_drag_2021}, which \textcolor{red}{is a departure from} mixture theory \citep{atkin1976continuum}. \textcolor{blue}{Key to such approaches for segregation velocity models is the closure relation} for pressure partitioning and inter-species drag. Despite progress in kinetic-theory-based segregation models for collisional granular flows \citep{jenkins_balance_1987,jenkins2002segregation,larcher2015evolution,neveu_particle_2022}, developing closures from first principles remains challenging for \textcolor{blue}{size-bidisperse dense granular flows, where multiple, enduring frictional contacts are common}. \textcolor{blue}{A variety of approaches and assumptions have been used to account for pressure partitioning and drag in dense granular flows \citep{gray2005theory,fan2011theory,marks2012grainsize,gajjar_asymmetric_2014,hill_segregation_2014}. Experiments have also been used to directly measure the segregation velocity of single intruder particles \textcolor{red}{and mixtures} undergoing shear~\citep{van_der_vaart_underlying_2015,trewhela_experimental_2021}. Alternatively, a variety of approaches using virtual springs tethered to a single intruder particle or groups of particles have been used to provide insight into the forces on particles~\citep{guillard_scaling_2016, jing_unified_2021, liu_lift_2021,bancroft_drag_2021, duan_segregation_2022, duan_no_segreg_2023, duan2024general}. In all cases, the challenge is to develop a general closure model for the segregation velocity that is applicable over a broad range of granular flow geometries.} 

In this paper, we develop \textcolor{blue}{such} a general closure model for the segregation velocity, exploiting recently established segregation force and drag force models at the particle level \citep{jing_rising_2020,jing_unified_2021,jing_drag_2022} along with their extensions to mixtures of arbitrary concentrations~\citep{duan_segregation_2022,duan2024general}. The approach follows the fluid terminal velocity analogue \textcolor{blue}{by using a force balance of the particle weight, the segregation force, and the granular drag force to determine the segregation velocity in mixtures of two particle species}. Since the segregation force and drag force models are characterized based on particle-resolved simulations (i.e., DEM simulations) with measurable parameters, they serve as first-principles closures for the species momentum balance equations at the continuum level and produce segregation velocity predictions matching simulation results for size-bidisperse granular flows across a diverse set of flow geometries.

\textcolor{blue}{\section{Equations of motion}}

\textcolor{blue}{Several models are combined to extract the segregation velocity. In this section, we outline these models and provide details on how to combine them to estimate the segregation velocity under a wide range of flow and particle conditions.}

\subsection{\textcolor{blue}{Particle force balance in a mixture}}

\begin{figure}\centerline{\includegraphics[width=0.8\columnwidth]{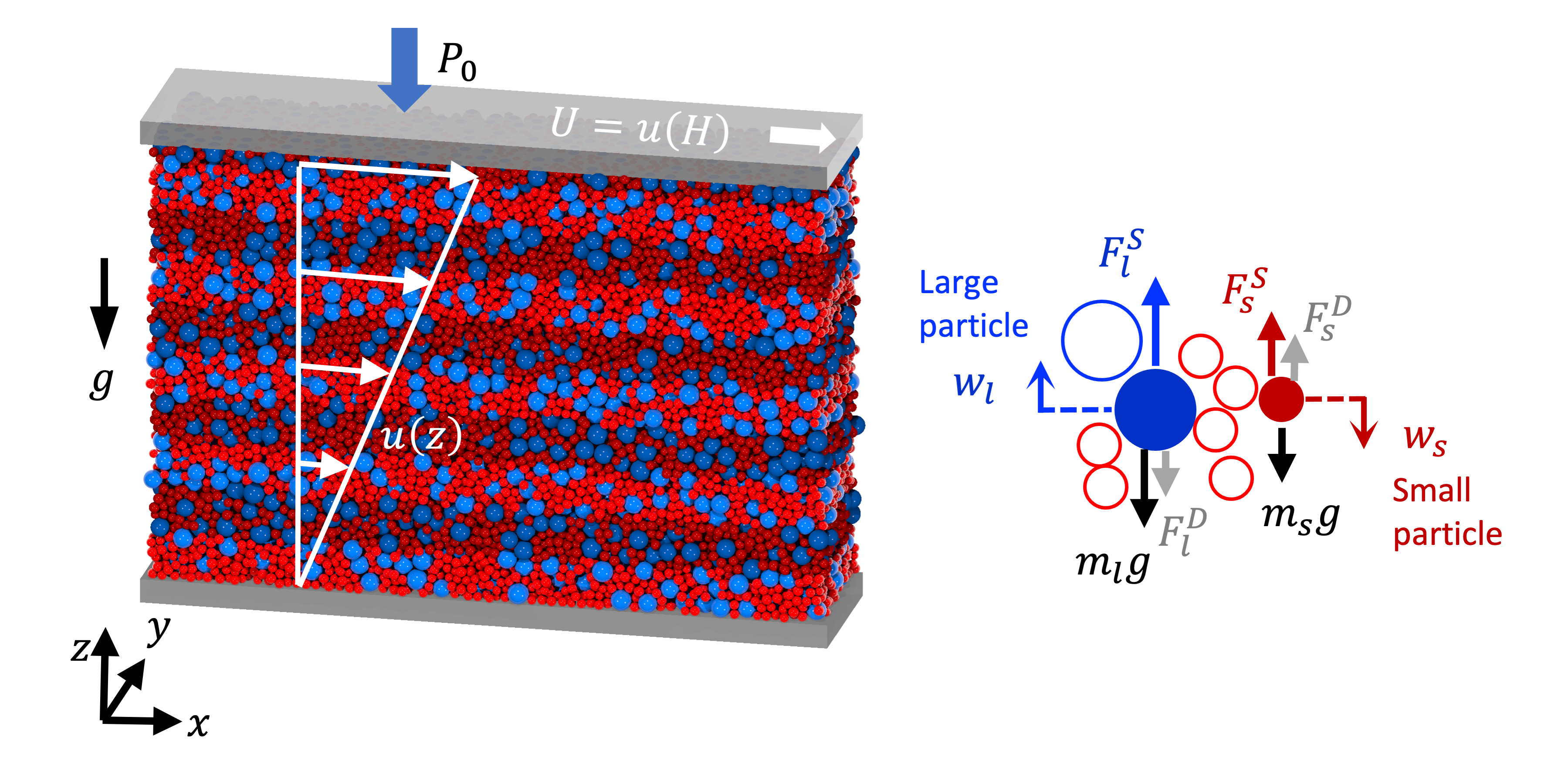}}
\caption{(left) DEM simulation example of large (4\,mm, blue) and small (2\,mm, red) spheres in a uniform shear flow with downward gravity (negative $z$-direction), partitioned into 2.5$d_l$ high layers (shading) for characterizing depth-varying segregation velocity. Here, large particles rise while small particle sink. The segregation direction varies in the different flow configurations analyzed later.
(right) Force balances on a large particle and a small particle corresponding to equation~(\ref{force_balance}) and \textcolor{blue}{species-specific vertical segregation velocities, $w_i$}.
}
\label{fig:schematics}
\end{figure}

\textcolor{blue}{Consider an incompressible flow of a size-disperse mixture of two species, such as the simple shear flow shown schematically in figure~\ref{fig:schematics}. The two particle species have volume concentrations $c_i$, where $i=l,s$ for large and small particles, respectively, and $c_s+c_l=1$. We neglect vertical acceleration terms, which is reasonable for the relatively slow segregation observed in many common granular flows including heap, chute, and rotating tumbler flows~\citep{gray2018particle, umbanhowar2019modeling}, though not necessarily all flows (such as some high-speed geophysical flows)}. 

\textcolor{blue}{A  force balance at the particle level in the vertical (gravitational, $g$) direction for an individual non-accelerating particle with mass $m_i$ in the flowing mixture, shown on the right in figure~\ref{fig:schematics}, includes the segregation force, $F^S_i$, the weight, $m_i g$, and the drag force, $F^D_i$ such that~\citep{jing_drag_2022}:}
\begin{equation}
        F^S_i-m_i  g+  F^D_i =0. 
\label{force_balance}
\end{equation}

\textcolor{blue}{Much like in the analysis of terminal velocity in a fluid, the segregation velocity in granular flows appears in the drag force, $F^D_i$, as we will show shortly. Thus, with an appropriate model for the segregation force, $F^S_i$, and a model for the dependence of the drag force, $F^D_i$, on the segregation velocity, we can use equation~(\ref{force_balance}) to calculate the segregation velocity. The key is having appropriate models for $F^S_i$ and $F^D_i$ on an individual particle in the flowing mixture, which are described shortly.}

\textcolor{blue}{Using force balance at the particle level, i.e.\ equation~(\ref{force_balance}), differs somewhat from the continuum description of segregation using the mixture theory framework. Within this framework, the momentum balance for each species along the segregation direction (negative $z$-direction) in a simple shear flow scenario can be expressed as \citep{gray2005theory, tunuguntla2017comparing}:
\begin{equation}
        -\frac{\partial p_i}{\partial z}-\rho_i  g+ {\beta}_i=0.
\label{momentum_simp}
\end{equation}
Here, $ -\partial p_i / \partial z= n_i F^S_i$ is the partial pressure gradient, where $p_i$ is the partial pressure of species $i$ and $z$ is in the direction of gravity, and $\rho_i$ is the density of species $i$. The interspecies momentum exchange is $\beta_i=n_i F^D_i$, where 
$n_i = c_i \phi/V_i$ represents the particle number density, $\phi$ is the bulk solid volume fraction, and $V_i$ denotes the individual particle volume of species $i$. 
Combined with the bulk pressure gradient $\partial p/\partial z = -\phi \rho g$, where it is assumed that both species have the same density $\rho$, the ratio of the pressure contribution of species $i$ to the bulk pressure $p$, or normal stress fraction, is $f_i=p_i/p= n_i F^S_i/\phi \rho g=c_i F^S_i/m_i g$ in the simplified case where $F^S_i$ remains constant with depth.
Prior studies adopting the momentum-based approach have often assumed a quadratic dependence or a particle-size or volume weighted dependence of $f_i$ on $c_i$~\textcolor{red}{\citep{marks2012grainsize, tunuguntla2014mixture, van_der_vaart_underlying_2015, gray2015particle, trewhela_experimental_2021, rousseau_bridging_2021}}, along with a linear drag model \citep{gray2005theory}. While these approaches have} \textcolor{red}{been proposed for} \textcolor{blue}{species concentration profiles in specific scenarios, the approach we use here matches} \textcolor{red}{direct DEM measurements of $f_i$} \textcolor{blue}{~\citep{duan_segregation_2022}, and using (\ref{force_balance}) allows us to consider the segregation velocity of not only mixtures but also intruders for a wide range of granular flow conditions and geometries.
}
\\
\subsection{Segregation force, \textcolor{blue}{$F^S_i$}}
Determining the segregation force $F^S_i$ at finite concentration starts with the segregation force on a single intruder, $F^S_{i,0}$ (subscript $0$ indicates the single intruder limit of species $i$, $c_i \to 0$). $F^S_{i,0}$ can be modelled with two additive terms, one related to gravity and the other to flow kinematics. \textcolor{blue}{Inspired by the observations of \citet{fan_phase_2011, fan2011theory}, the flow kinematics term was initially scaled with the shear stress gradient \citep{guillard_scaling_2016} but was later linked to the shear rate gradient \citep{jing_unified_2021, singh2024continuum}. The  segregation force can be expressed as \citep{jing_unified_2021}}:
\begin{equation}
    \label{eq:intruder}
    F^S_{i,0}=-f^g(R_d)\frac{\partial{p}}{\partial{z}}V_i+f^k(R_d)\frac{p}{\dot\gamma}\frac{\partial\dot\gamma}{\partial{z}}V_i,
\end{equation}
where superscripts $g$ and $k$ indicate gravity- and kinematics-related mechanisms, respectively, $V_i$ is the intruder particle volume, $\dot\gamma$ is the local shear rate, and $\rho$ is the density of both the intruder and the bed particles. 
The gravity term is buoyancy-like, and the kinematic term depends on the shear rate gradient in the flow.
The empirical dimensionless functions $f^g(R_d)$ and $f^k(R_d)$ depend on the intruder-to-bed-particle diameter ratio $R_d=d_i/d_j$ \citep{jing_unified_2021},
\begin{subequations}
\begin{equation}
        f^g(R_d)=\bigg[ 1-c^g_1 \exp(-\frac{R_d}{R^g_1}) \bigg] \bigg[ 1+c^g_2\exp(-\frac{R_d}{R^g_2}) \bigg],
\end{equation}
\begin{equation}
        f^k(R)=f^k_\infty \bigg[ \tanh{(\frac{R_d-1}{R^k_1})} \bigg] \bigg[ 1+c^k_2\exp(-\frac{R_d}{R^k_2}) \bigg],
\end{equation}
\label{eq:intruder_coefficient}
\end{subequations}\\
where $R^g_1=0.92$, $R^g_2=2.94$, $c^g_1=1.43$, $c^g_2=3.55$, $f^k_\infty=0.19$, $R^k_1=0.59$, $R^k_2=5.48$, and $c^k_2=3.63$ are fitting parameters appropriate for a range of flow conditions. In applying these functions over a range of concentrations, we need to consider both a large intruder particle surrounded by a bed of small particles and a small intruder particle surrounded by a bed of large particles corresponding to intruder-to-bed particle size ratios of $d_l/d_s$ and $d_s/d_l$, respectively. 


To predict the segregation force $F^S_i$ \textcolor{blue}{in mixtures} at arbitrary non-zero concentrations, the intruder segregation force model in Eq.~(\ref{eq:intruder}) is extended using semi-empirical relations \textcolor{blue}{for mixtures (rather than an intruder particle)} based on DEM simulations \citep{duan_segregation_2022}. For large particles in a bidisperse mixture of concentration $c_l$~\citep{duan2024general},
\begin{subequations}
\textcolor{blue}
{
\begin{equation}
 F^S_l=m_l g\cos{\theta}+(F_{l,0}^S-m_l g\cos{\theta})\textrm{tanh}\Big( \frac{m_l g\cos{\theta}- F_{s,0}^S }{F_{l,0}^S-m_l g\cos{\theta}}\frac{c_s}{c_l} \Big),
\label{tanh_large2}
\end{equation}}
and for small particles,
\textcolor{blue}
{\begin{equation}
    F_{s}^S=m_s g\cos{\theta}-(F_{l,0}^S-m_s g\cos{\theta}){\frac{c_l}{c_s}}\textrm{tanh}\Big( \frac{m_s g\cos{\theta}-F_{s,0}^S}{{F}_{l,0}-m_s g\cos{\theta}}\frac{c_s}{c_l} \Big),
\label{tanh_small2}
\end{equation}
\label{eq:mixture_tanh1}}
\end{subequations}\\
where $\theta$ is the angle between gravity and the segregation direction, which is generally normal to the flow.

Equations~(\ref{eq:mixture_tanh1}) can be rewritten in terms of the net forces ($T_i=F^S_i-m_ig\cos\theta$) that balance the inter-species drag (\ref{force_balance}) such that, 
\begin{subequations}
\begin{equation}
 T_l=T_{l,0}\textrm{tanh}\Big( -\frac{{T}_{s,0}}{{T}_{l,0} }\frac{m_lc_s}{m_sc_l} \Big),
\label{tanh_large}
\end{equation}
\begin{equation}
    T_{s}=-{T}_{l,0}{\frac{m_sc_l}{m_lc_s}}\textrm{tanh}\Big( -\frac{{T}_{s,0}}{{T}_{l,0} }\frac{m_lc_s}{m_sc_l} \Big),
\label{tanh_small}
\end{equation}
\label{eq:mixture_tanh}
\end{subequations}\\
where $T_{i,0}=F_{i,0}-m_ig\cos \theta$, and $F_{i,0}$ and $F_{i}$ denote the segregation forces for a single intruder and for mixtures, respectively. 
The force balance in~(\ref{force_balance}) can be rewritten \textcolor{blue}{for a mixture} as
\begin{equation}
        T_i+  F^D_i =0. 
\label{eq:force_balance_t}
\end{equation}
\textcolor{blue}{What remains to be specified is an expression for the drag force in a mixture, $F^D_i$, including its dependence on the segregation velocity, which is described next.}

\subsection{Drag \textcolor{blue}{force, $F^D_i$}}
\label{section:drag}
\textcolor{blue}{As with the approach for the segregation force, we start with the drag force on a single intruder particle within a monodisperse flow of bed particles, $F^D_{i,0}$. A} Stokes drag formulation can be employed to express the drag force \citep{tripathi2013density,jing_drag_2022,he2025lift} \textcolor{blue}{in terms of the intruder segregation velocity relative to the local bulk flow velocity in the segregation direction, $w_{i,0}$,} as
\begin{equation}
    \label{eq:drag}
    F^D_{i,0}=-C^D_{i,0} \pi \eta d_i w_{i,0},
\end{equation}
where $C^D_{i,0} $ is the drag coefficient for a single intruder, and $\eta$ is the \textcolor{blue}{effective bulk granular viscosity calculated from the $\mu(I)$ rheology as described shortly}.
For Stokes drag on a spherical particle at low Reynolds number in a viscous fluid, $C^D_{i,0}=3$. However, the value of $C^D_{i,0}$ for an intruder in a granular flow is not as simply specified.
For a single spherical intruder particle, $C^D_{i,0} \approx 2.1$ for $1\leq R_d \leq 5$, but the precise value depends on the flow conditions~\citep{jing_drag_2022}\textcolor{blue}{, specifically, the size-bidispere mixture inertial number~\citep{rognon_dense_2007, tripathi_rheology_2011}}, 
\begin{equation}
    \label{eq:mix_I}
   \textcolor{blue}{ I =\dot\gamma \sqrt{\frac{\rho}{p}}\sum_{i=s,l}d_ic_i.} 
\end{equation}
For large intruders with $R_d\ge1$,
\begin{equation}
    \label{eq:drag_coefficient}
    C^D_{i,0} =[k_1-k_2\exp(-k_3 R_d)]+s_1 I R_d +s_2 I (R_\rho-1), 
\end{equation}
where $k_1=2$, $k_2=7$, $k_3=2.6$, $s_1=0.57$, and $s_2=0.1$ are fitting parameters determined across a wide range of flow conditions ($0.6\leq R_d \leq 5$, $1\leq R_{\rho} \leq 20$, and $I \lesssim 1$), \textcolor{blue}{and $R_\rho$ is the intruder-to-bed-particle density ratio}~\citep{jing_drag_2022}.
The granular viscosity $\eta$ is estimated from the $\mu(I)$ rheology \citep{midi2004dense} as
\begin{subequations}
\begin{equation}
    \label{eq:mu}
    \eta=\mu_\mathrm{eff} \frac{p}{\dot\gamma},
\end{equation}
where the effective friction coefficient is
\begin{equation}
    \label{eq:mu_eff}
    \mu_\mathrm{eff}=\tau/p=\mu_s+\frac{\mu_2-\mu_s}{I_c/I+1},
\end{equation}
\label{eq:viscosity}
\end{subequations}\\
where $\tau$ is the shear stress and $\mu_s$, $\mu_2$, and $I_c$ are granular material specific parameters. \textcolor{blue}{The flows simulated in this study generally follow the $\mu(I)$ rheology} \textcolor{red}{for dense flows with the usual caveats for collisional flow  at larger $I$ than considered here and for quasi-static flow as $I\rightarrow 0$}, see appendix~\ref{appendixA}.

The drag \textcolor{blue}{coefficient, $C^D_{i,0}$, in (\ref{eq:drag_coefficient}) is} for a single intruder particle in an otherwise homogeneous bed of the other species. Since we are interested in the segregation velocity of particles in a mixture, it is necessary to determine the dependence of the drag force in a mixture, $F^D_i,$ on species concentration. Previous approaches to determine \textcolor{blue}{the mixture} $C^D_i$ have predominantly focused on density-bidisperse mixtures where $R_d=1$ \citep{tripathi2013density,duan2020segregation,bancroft_drag_2021} 
due to the simplicity of estimating the segregation force in terms of buoyancy. Reported values of $C^D_i$ at $R_d=1$ based on this approach range from 1.7 to 3.7 depending on the volume fraction, and $C^D_i$ is independent of $c_i$  for density-bidisperse mixtures. However, the concentration dependence of $C^D_i$ in size-bidisperse mixtures ($R_d\ne 1$ and density ratio $R_\rho=1$) has not been considered explicitly, \textcolor{blue}{although~\cite{bancroft_drag_2021} mention it in passing. We use simulations of controlled shear flow later in this paper (section~\ref{sec:Drag_mix}) to demonstrate that the drag coefficient is nearly independent of the mixture concentration for the conditions we consider here. For now, in order to proceed with the analysis, we simply assume that $C^D_i \approx C^D_{i,0}$. Hence,}
\textcolor{blue}
{\begin{equation}
    F^D_i \approx F^D_{i,0}=-C^D_i \pi \eta d_i w_i,  
\label{eq:drag_approx}
\end{equation}
where the mixture drag coefficient, $C^D_i$, has been substituted for the intruder drag coefficient, $C^D_{i,0}$, and the species-specific segregation velocity in the mixture, $w_i$, for the intruder segregation velocity, $w_{i,0}$, in (\ref{eq:drag}). }

\subsection{Segregation velocity}
\label{section:seg_vel}
\textcolor{blue}{The species-specific segregation velocity, $w_i$, relative to the local bulk flow velocity in the segregation direction is now easily calculated by substituting the mixture drag force (\ref{eq:drag_approx}) and the segregation force model (\ref{eq:mixture_tanh}) into the force balance (\ref{eq:force_balance_t}) and solving for the segregation velocities of the large particle species, $w_l$, and small particle species, $w_s$:}
\begin{subequations}
\begin{equation}
 w_l=\frac{ T_{l,0}\textrm{tanh}\Big( -\frac{{T}_{s,0}}{{T}_{l,0} }\frac{m_lc_s}{m_sc_l} \Big)}{C^D_l \pi \eta d_l}
\label{eq:wl}
\end{equation}
and
\begin{equation}
 w_s=-\frac{ {T}_{l,0}{\frac{m_sc_l}{m_lc_s}}\textrm{tanh}\Big( -\frac{{T}_{s,0}}{{T}_{l,0} }\frac{m_lc_s}{m_sc_l} \Big) }{C^D_s \pi \eta d_s}.
\label{eq:ws}
\end{equation}
\label{eq:w}
\end{subequations}\\
\textcolor{blue}{Recall that} we assume $C^D_i \approx C^D_{i,0}$, independent of species concentration $c_i$, \textcolor{blue}{which we confirm below in section~\ref{sec:Drag_mix}.}


\subsection{\textcolor{blue}{Effect of diffusion on segregation velocity}}
\label{section:diff}
\textcolor{blue}{A final consideration is the diffusive flux of species driven by collisional diffusion and its effect on the measured segregation velocity. The diffusion contribution is most clearly framed in terms of the} advection-diffusion-segregation transport equation based on mass balance \textcolor{blue}{that} has been successfully used to model segregation in flowing granular media \citep{bridgwater1985particle,dolgunin1995segregation,gray2018particle,umbanhowar2019modeling}. 
Within this continuum framework, the concentration of species $i$ can be expressed as 
\begin{equation}
\frac{\partial c_i}{\partial t} + {\div (\pmb u_i c_i)}={\div (D\nabla c_i)}.
\label{transport}
\end{equation}
Here, \textcolor{blue}{$\pmb u_i$ is the diffusionless velocity, and} the local collisional diffusion coefficient $D$ is a scalar, although in general it is a tensor. This approximation is accurate for flows with a single dominant shear direction \citep{umbanhowar2019modeling}. Note that the diffusionless velocity, $\pmb u_i$, differs slightly from the overall velocity, which is a combined effect of both advection and diffusion. 
\textcolor{blue}{With the usual assumptions of two-dimensional flow and gradual development in the streamwise direction}, equation~(\ref{transport}) in the $z$-direction can be written as
\begin{equation}
\frac{\partial c_i}{\partial t} +\frac{\partial  (w_i+w)c_i}{\partial z}=\frac{\partial}{\partial z} \Big( D\frac{\partial c_i}{\partial z} \Big),
\label{transport1}
\end{equation}
or, rearranging, as
\begin{equation}
\frac{\partial c_i}{\partial t} +\frac{\partial  \big[(w_i+w)c_i-D(\partial c_i/\partial z) \big] }{\partial z}=0,
\label{transport2}
\end{equation}
\textcolor{blue}{where $w$ is the local vertical velocity of the bulk. Note that $w=0$ in the reference frames associated with the example flows used in this study}.

When the normal component of flux for species $i$ is measured from DEM simulation, it is the entire quantity within the bracket of (\ref{transport2}) that is measured. 
In other words, the measured normal flux $(w_i+w)c_i-D(\partial c_i/\partial z)$ is a combination of both the \textcolor{blue}{total segregation} flux, ($w_i+w)c_i$, and the diffusion flux, $-D(\partial c_i/\partial z)$.
To compare the segregation velocity model predictions developed in this paper with DEM measurements of the segregation velocity, the \textcolor{blue}{segregation} flux needs to be combined with the diffusion flux, such that the net \textcolor{blue}{species} velocity is
\begin{equation}
    w^{net}_i=w_i-\frac{D}{c_i}\frac{\partial c_i}{\partial z},
    \label{eq:w_net}
\end{equation}
which can be directly measured in situations where there is a concentration gradient.



Both experimental \citep{bridgwater1980self,utter2004self} and
computational \citep{fan2015shear,cai2019diffusion,fry2019diffusion} studies of dense granular flows suggest that the diffusion coefficient, $D$, is proportional to the product of the local shear rate and the square of the local mean particle diameter,
\begin{equation}
D=A \dot\gamma \bar d ^2,
\label{diffusionmodel}
\end{equation}
where $\bar d = \sum c_i d_i$ and $A$ is a constant with reported values in the range 0.01 to 0.1 \citep{savage1993studies,hsiau1999fluctuations,utter2004self,fan2014modelling,fan2015shear,cai2019diffusion,fry2019diffusion}.
In this study, $A=0.046$ based on diffusion coefficient data measured from heap flow simulations \citep{duan_segregation_2022}.

\textcolor{blue}{With the exception of demonstrating that the drag coefficient for a mixture is similar to that for an intruder particle, as noted in section~\ref{section:drag}, we now have all of the relationships necessary to calculate the segregation velocity. Before addressing drag in mixtures, it is first necessary to describe the simulation approach that we use.}
\\
\section{Simulations}

An in-house discrete element method (DEM) code running on CUDA-enabled NVIDIA GPUs is used to simulate size-bidisperse particle mixtures with species specific volume concentration $c_i$, diameter $d_i$, and density $\rho_l=\rho_s=1$\,g\,cm$^{-3}$.  
Large ($d_l=4$\,mm) and small ($d_s$ varied to adjust the size ratio, $R_d=d_l/d_s$) particle species have a $\pm10$\% uniform size distribution to minimize layering~\citep{staron_segregation_2014} (increasing the diameter variation to $\pm20$\% does not alter the results). The mixture is sheared in the streamwise ($x$) direction (see figure~\ref{fig:schematics}).
Boundary conditions are periodic in $x$ and $y$ with length $L=35d_l$ and width $W=10d_l$, respectively. The height is $H=50d_l$ in the $z$-direction, which is normal to the flow direction (reducing $H$ to $25d_l$ does not alter the results).
In all cases, particles fall freely under gravity to fill the domain before flow begins.  Gravity may be aligned with the $z$-direction, as shown in figure~\ref{fig:schematics}, at an angle $\theta$ with respect to $z$ for inclined chute flow, or parallel to the flow aligned with $x$ for vertical chute flow. In some cases, gravity is set to zero; in all other cases we use $g=g_0\equiv 9.81$\,m\,s$^{-2}$. 

The standard linear spring-dashpot model~\citep{cundall1979discrete} is used to resolve particle-particle and particle-wall contacts of spherical particles using a friction coefficient of $\mu=0.5$, a restitution coefficient of 0.9, and a binary collision time of 0.15\,ms. We have confirmed that our results are relatively insensitive to these values except for very low friction coefficients ($\mu\le0.2$)~\citep{duan2020segregation,jing_rising_2020}.
From 26000 to 150000 particles are included in each simulation, depending on the size ratio.

The segregation velocity is measured in a variety of flow configurations, including controlled shear flows and natural uncontrolled flows. These various flow configurations are explained in more detail in a previous paper in which we consider the segregation force \citep{duan2024general}.
The first flow conditions that we consider are controlled shear flows in which the velocity profile is constrained to be of a certain form. The specified velocity profile, $u(z)$, is achieved by applying a small streamwise stabilizing force $k_v [\,u(z)-u_p(z_p)]\,$ to each particle at each DEM simulation time step in order to maintain the desired velocity profile, where $u_p$ and $z_p$ are the instantaneous particle velocity and position, respectively, and $k_v$ is a gain parameter~\citep{lerner_unified_2012, clark_critical_2018, fry2018effect, jing_rising_2020,jing_unified_2021, jing_drag_2022}.
By prescribing a specific velocity profile, we control the shear rate and shear rate gradient, which play direct roles in determining both the segregation force (\ref{eq:intruder}) and the drag (\ref{eq:drag}) via the viscosity (\ref{eq:viscosity}), and hence influence the segregation velocity (\ref{eq:w}). The presence of gravity results in a pressure gradient in $z$, which also influences the segregation force (\ref{eq:intruder}).
We consider four cases: a linear velocity profile with gravity, an exponential velocity profile without gravity, a parabolic velocity profile without gravity, and an exponential velocity profile with gravity. \textcolor{blue}{(For reference, each of these flows is shown schematically later in the paper as insets when the results are discussed, see figure~\ref{fig:controlled}(a).) A wide range of inertial numbers, $I$, are achieved via the variation of the pressure with depth for flows with $g\neq 0$ and by imposing large shear rates, which can lead to wall velocities of $u(H)=U=20$\,m\,s$^{-1}$ in some cases. }
For the flows with gravity (linear and exponential velocity profiles),  a small overburden pressure $P_0$ equal to the pressure at
a depth of $2.5 d_l$ (i.e., $P_0=0.05\rho \phi g_0 H$, where the bulk solid fraction $\phi\approx 0.55$) is imposed on the upper wall, which is free to move vertically, and which fluctuates in height by no more than ±0.05\% after an initial rapid dilatation of the particles at flow onset. 
For the flows without gravity (exponential and parabolic velocity profiles), the top wall is fixed vertically. 
These different velocity profiles allow us to consider cases with no pressure gradient (when gravity $g=0$) so that the gravity-related first term of (\ref{eq:intruder}) is zero, with no shear rate gradient (linear profile) so that the kinematics-related second term of (\ref{eq:intruder}) is zero, and with combinations of the gravity and shear such that both the gravity and kinematics terms in (\ref{eq:intruder}) contribute to the segregation velocity.

In addition to the controlled shear flows, we also consider four cases where the velocity field is not directly controlled. \textcolor{blue}{(For reference, each of these flows are shown schematically later in the paper as insets when the results are discussed, see figure~\ref{fig:natural}.)}  The flow kinematics of these uncontrolled ``natural flows" are driven entirely by the combined effects of gravity and boundary conditions. The walls are rough in all cases, formed from a $2.5d_l$ thick layer of bonded large and small particles that move collectively. For the wall-driven flows, an overburden pressure $P_0$ equal to the pressure at
a depth of $H/2$ (i.e. $P_0=0.5\rho \phi g_0 H$) is imposed on the upper wall.
When gravity is included for plane shear flow and inclined chute flow, it results in a pressure gradient in $z$. In both wall driven flows, the upper wall moves at velocity $u(H) = 10\,$m\,s$^{-1}$ in the $x$-direction and the lower wall at $u(0) = -10\,$m\,s$^{-1}$ in the negative $x$-direction. Both cases show little to no slip at either wall.  The vertical chute flow is driven by gravity, which is aligned parallel to the rough fixed bounding walls, resulting in a generally uniform velocity at the centre of the channel that goes to zero at the walls. In this case, there is no pressure gradient in $z$ to drive segregation, so any segregation in $z$ is driven by shear gradients alone. Finally, the inclined chute flow lacks an upper wall (free boundary) so that particles flow due to a streamwise component of gravity. Here the pressure gradient in the segregation direction is $g_0 \cos{\theta}$, where $\theta$ is the inclination angle of the base (lower wall) relative to $\vec{g}$.

To consider segregation for each of these flow conditions, the simulation domain is discretized into horizontal layers, each of thickness $2.5d_l$ (1\,mm) in the $z$-direction, for averaging purposes (see figure~\ref{fig:schematics}). 
Decreasing the layer thickness to $1.25d_l$ increases averaging uncertainties but does not alter the mean values of the flow fields.
Within each layer, various local variables are measured including the streamwise velocity ($u$),  pressure ($p$), shear rate ($\dot \gamma$), and species concentration ($c_i$). Subsequently, these flow measurements, which are averaged in the $x-$ and $y-$directions but vary in the $z-$direction, are used to determine intermediate variables including the net gravity and segregation force acting on each species ($T_i$ via (\ref{eq:intruder}-\ref{eq:mixture_tanh})), the local viscosity of the mixture ($\eta$ via (\ref{eq:viscosity})), the drag coefficient for each species ($C^D_i$ via (\ref{eq:drag_coefficient})), and the diffusion coefficient ($D$ via (\ref{diffusionmodel})).
Finally, these computed variables are used to predict the segregation velocity using equations (\ref{eq:w}) and, \textcolor{blue}{where necessary,} (\ref{eq:w_net}).

The predicted segregation velocities based on the model of (\ref{eq:w}) are compared to the segregation velocities measured from the simulations. To characterize the segregation velocity for each layer in figure~\ref{fig:schematics}, we assess the average center of mass height for each species relative to the mean height of all particles over a short measurement window, calculated as:
\begin{equation}
    \bar z_i = \frac{1}{N_i}\sum^{N_i}_{k\in i}z_k- \frac{1}{N}\sum^{N}_{k=1}z_k,
    \label{eq:com}
\end{equation}
where $N_i$ and $N$ are the number of particles of species $i$ and the total number of particles in the horizontal averaging layer, respectively.
The segregation distance \textcolor{blue}{for species \textit{i} is the offset of its} center of mass from its initial position, $\bar z_i - \bar z_{i,0}$. The segregation velocity \textcolor{blue}{for species \textit{i} is then} measured as the rate of this offset, $(\bar z_i - \bar z_{i,0})/\Delta t$, where the measurement window is $\Delta t=1\,s$, which we have shown previously is sufficiently long to provide statistically meaningful data and short enough to capture temporally local results \citep{duan2020segregation}.

To mitigate the influence of noise and kinematic acceleration on the measurements, \textcolor{blue}{each simulation begins with the initial flow of mixed particles subject to equal and opposite vertical restoring forces applied to particles of each species in each layer in figure~\ref{fig:schematics}. This technique maintains the initial uniformly mixed distribution of the two particle species and suppresses segregation while the flow develops, similar to the approach previously employed to measure the mixture segregation force \citep{duan_segregation_2022, duan2024general}.
The flow is allowed to develop for 2\,s} with the prescribed velocity and concentration profiles and the segregation being suppressed. During the subsequent 1\,s measurement window, the restoring forces suppressing the segregation are deactivated while primary flow field parameters, such as streamwise velocity, pressure, and \textcolor{blue}{particle vertical positions ($z_k$)} are recorded at intervals of 0.01\,s. The \textcolor{blue}{species segregation} velocity is then measured as $w_i=(\bar z_i-\bar z_{i,0})/ \Delta t$ for the ensemble of \textcolor{blue}{each particle species} in each layer, and the other local variables ($u$, $p$, $c_i$) in each layer are averaged over the 1\,s measurement window for use in calculating the predicted segregation velocity from (\ref{eq:w}).
\textcolor{blue}{
We have shown previously~\citep{fry2018effect, duan2020segregation, duan_segregation_2022, duan2024general} and confirmed here that the velocity profile is unaffected by the segregation for the short duration of the measurement window. Furthermore, we have confirmed that the concentration profile in the bulk changes by less than 2\% on average over the 1\,s measurement window.
}

\section{\textcolor{blue}{Drag force in mixtures}}
\label{sec:Drag_mix}
 
Before considering the segregation velocity, which is the focus of this paper, it is necessary to address \textcolor{blue}{the effect of mixture concentration on the drag force, as noted in section \ref{section:drag}.} 
To extend the intruder drag model of equations~(\ref{eq:drag}) and (\ref{eq:drag_coefficient}) to mixtures, we perform a series of simulations using an approach \textcolor{blue}{motivated by \cite{bancroft_drag_2021},} where opposing forces are applied to particles of each species. Specifically, equal and opposite applied forces are imposed in the segregation direction to particles of each species (negative $z$-direction for small particles and positive $z$-direction for large particles) in a homogeneous shear flow of a mixture of large and small particles like that shown in figure~\ref{fig:schematics} with $g=0$. The applied force, which is distributed across all particles of a species, drives the particle species to segregate at a rate controlled entirely by the applied force and the drag, which balance each other\textcolor{blue}{~\citep{jing_drag_2022}. In this way, (\ref{force_balance}) has just two terms, the mixture segregation force, $F^S_i$, which is equivalent to the applied force, and the mixture drag force $F^D_i$ (since $g=0$). $F^D_i$ is given by (\ref{eq:drag_approx}) which is in terms of the effective granular viscosity, $\eta$, and} the species-specific segregation velocity, $w_i$. By tracking the average motion of all of the particles of each species, $w_i$ for species $i$ is obtained. By calculating the overall normal and shear stresses, $P$ and $\tau$, from inter-particle collisions \citep{luding2008introduction}, $\eta$ is obtained via~(\ref{eq:viscosity}). \textcolor{blue}{Using force balance (\ref{force_balance}) with the applied force for $F^S_i$, the mixture} drag coefficient $C^D_i$ is estimated using  \textcolor{blue}{(\ref{eq:drag_approx}) for $F^D_i$ with the measured $w_i$ and calculated $\eta$}. In these simulations, the domain is the same as the controlled shear flows used for measuring segregation velocities.  The data are recorded at intervals of 0.01\,s over a 1\,s window after the flow reaches a steady state. 

\begin{figure}\centerline{\includegraphics[width=\columnwidth]{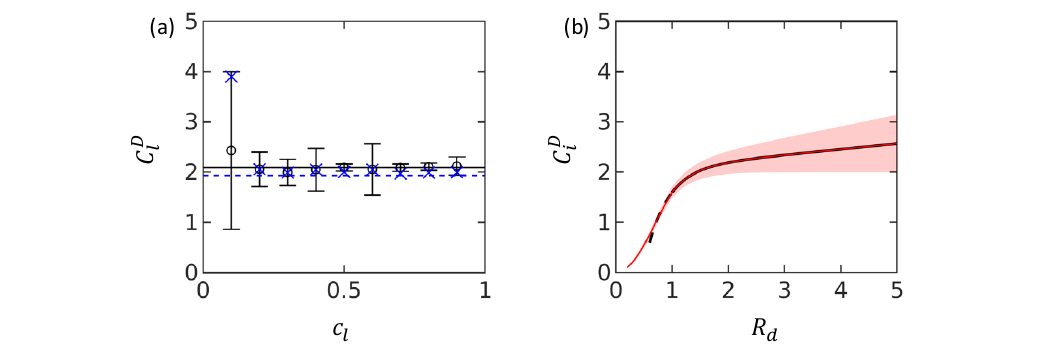}}
\caption{
(a) 
\textcolor{blue}{
Large particle drag coefficient, $C^D_l$, vs.\ large particle species concentration, $c_l$, in a uniformly sheared flow for size ratios of $R_d=1.5$ at $I\approx 0.08$ (blue crosses) and $R_d=2$ at $I\approx 0.12$ (black circles) for $g=0$. 
Error bars show the standard deviation of $C^D_{l}$ over a 1\,s window for $R_d=2$; error bars for $R_d=1.5$ are similar but omitted for clarity.  Horizontal solid black line corresponds to $C^D_{i,0}$ for $R_d=2$; horizontal dashed blue line corresponds to $C^D_{i,0}$ for $R_d=1.5$.
}
(b) \textcolor{blue}{Comparison of $C^D_{i,0}$ with $C^D_i$ for varying size ratio. The single intruder drag coefficient, $C^D_{i,0}$ is calculated from}~(\ref{eq:drag_coefficient}) for large ($i=l$ for $R_d\ge1$) (solid black curve) and small ($i=s$ for $R_d<1$) (dashed black curve) intruder particles. \textcolor{blue}{The mixture drag coefficient, $C^D_i$, (red curve) is calculated from (\ref{eq:drag_coefficient}) for $R_d\ge1$ and (\ref{eq:conservation}) for $R_d<1$.} Both curves represent predictions for $I=0.2$. \textcolor{blue}{Predictions of the mixture model for $I$ values ranging from 0 (lower bound) to 0.4 (upper bound), which are typical of dense granular flows, are indicated by the shaded band.}}
\label{fig:drag_coefficient}
\end{figure}

Figure~\ref{fig:drag_coefficient}(a) shows the drag coefficient of large particles measured in mixtures of particles with varying large particle concentration, $c_l$, in uniform shear flows with \textcolor{blue}{$R_d=1.5$ and 2, having a constant inertial number in each case.}
The values for $C^D_l$ are nearly independent of $c_l$ \textcolor{blue}{except at very low $c_l$, approaching the single intruder limit. Furthermore, the values for $C^D_l$}  match the value for $C^D_{l,0}$ (horizontal \textcolor{blue}{dashed line for $R=1.5$ and solid line for $R=2$}). Hence, the intruder drag model (\ref{eq:drag}) for $C^D_{l,0}$ provides a reasonable estimate of  $C^D_l$ for  mixtures at arbitrary non-zero concentrations \textcolor{blue}{for the cases considered here, although further study of the dependence of drag on species concentration, size ratio, and inertial number is warranted}.
 
Unlike the intruder drag model (\ref{eq:drag}) that treats the drag force on the intruder, whether it is small or large, as if it is in a sea of particles of the other size, the drag forces for a mixture must also satisfy volume conservation when determining the \textcolor{blue}{species-specific segregation velocities, $w_i$, for a mixture}.
This requires that the overall drag force for all of the large particles at a given concentration of large particles must be balanced by the drag force for all of the small particles at the corresponding concentration of small particles, while at the same time volume flux must be conserved \textcolor{blue}{(in the laboratory reference frame}), 
\begin{equation}
   w_l c_l+w_s c_s=0.
    \label{eq:vol_flux}
\end{equation}
Consequently, there exists an implicit relation between the drag coefficients for large ($C^D_l$) and small ($C^D_s$) particles to assure that volume flux is conserved. Noting from (\ref{eq:force_balance_t}) that $T_i+F^D_i=0$, expressions for $w_i$ for each species from (\ref{eq:drag_approx}) can be substituted into (\ref{eq:vol_flux}) yielding
\begin{equation}
    \frac{T_l}{C^D_l d_l} c_l + \frac{T_s}{C^D_s d_s} c_s =0.
\end{equation}
Using (\ref{eq:mixture_tanh}), this can be expressed as
\begin{equation}
    \frac{T_{l,0}}{C^D_l d_l} c_l - \frac{T_{l,0}}{C^D_s d_s} c_s \frac{m_sc_l}{m_lc_s} =0,
\end{equation}
which can be rearranged as
\begin{equation}
    \frac{C^D_l}{C^D_s}=\frac{m_ld_s}{m_sd_l}=R_\rho R_d^2.
    \label{eq:conservation}
\end{equation}
To satisfy this constraint for size segregation with density ratio $R_\rho=1$, we implement the drag model so that the large particle drag coefficient $C^D_l$ is estimated by equation~(\ref{eq:drag_coefficient}) but the small particle drag coefficient $C^D_s$ is calculated based on the correlation (\ref{eq:conservation}). 
The justification for this approach is that (\ref{eq:drag_coefficient}) is valid for $0.6\le R_d \le 5$, so we only use it to estimate $C^D_l$ for large particles ($R_d\ge1$) and use (\ref{eq:conservation}) to find the corresponding value for $C^D_s$ for small particles that assures volume flux conservation.
Figure~\ref{fig:drag_coefficient}(b) compares $C^D_i$ predictions by both the original intruder drag model (\ref{eq:drag_coefficient}) and the revised mixture drag model ((\ref{eq:drag_coefficient}) for $R_d\ge1$ and (\ref{eq:conservation}) for $R_d<1$) for $I=0.2$. 
The two approaches overlap for $R_d>0.6$, but the revised model (red curve) allows calculation of $C^D_s$ for \textcolor{blue}{$0.3 \lesssim R_d<0.6$}, while simultaneously assuring that volume flux is conserved.
Note that $C^D_i$ is nearly independent of $I$ for $R_d<1.5$. Outside this range, $C^D_i$ depends on $I$, with the band of values in figure~\ref{fig:drag_coefficient}(b) corresponding to $0\le I\le0.4$. However,  the effect of $C^D_i$ on the drag is relatively limited  compared to the influence of viscosity due to the constrained range of variation in $C^D_i$. \textcolor{blue}{Further note that the lower limit in $R_d$ below which (\ref{eq:conservation}) should not be applied is necessary because sufficiently small particles can freely pass through interstices between large particles even if the large particles are not flowing. We estimate this lower limit to be $R_d \approx 0.3$ based on studies of the segregation velocity of small particles in sheared beds~\citep{gao2024shearpercolation}.}

\section{Segregation Velocity Results}

In this section, we address the central result of this paper.  That is, we combine the models for the segregation force (\ref{eq:mixture_tanh}) in mixtures and the drag force for single intruders (\ref{eq:drag}) adapted to mixtures (\ref{eq:drag_approx}) via a simple \textcolor{blue}{mixture} force balance (\ref{eq:force_balance_t}) to predict the local segregation velocity via (\ref{eq:w}).  We then compare the predicted segregation velocity to the result directly measured from the DEM simulated flow to demonstrate the validity of the force-based modelling approach.

\subsection{Controlled shear flows}
\label{controlled}

\begin{figure} \centerline{\includegraphics[width=\columnwidth]{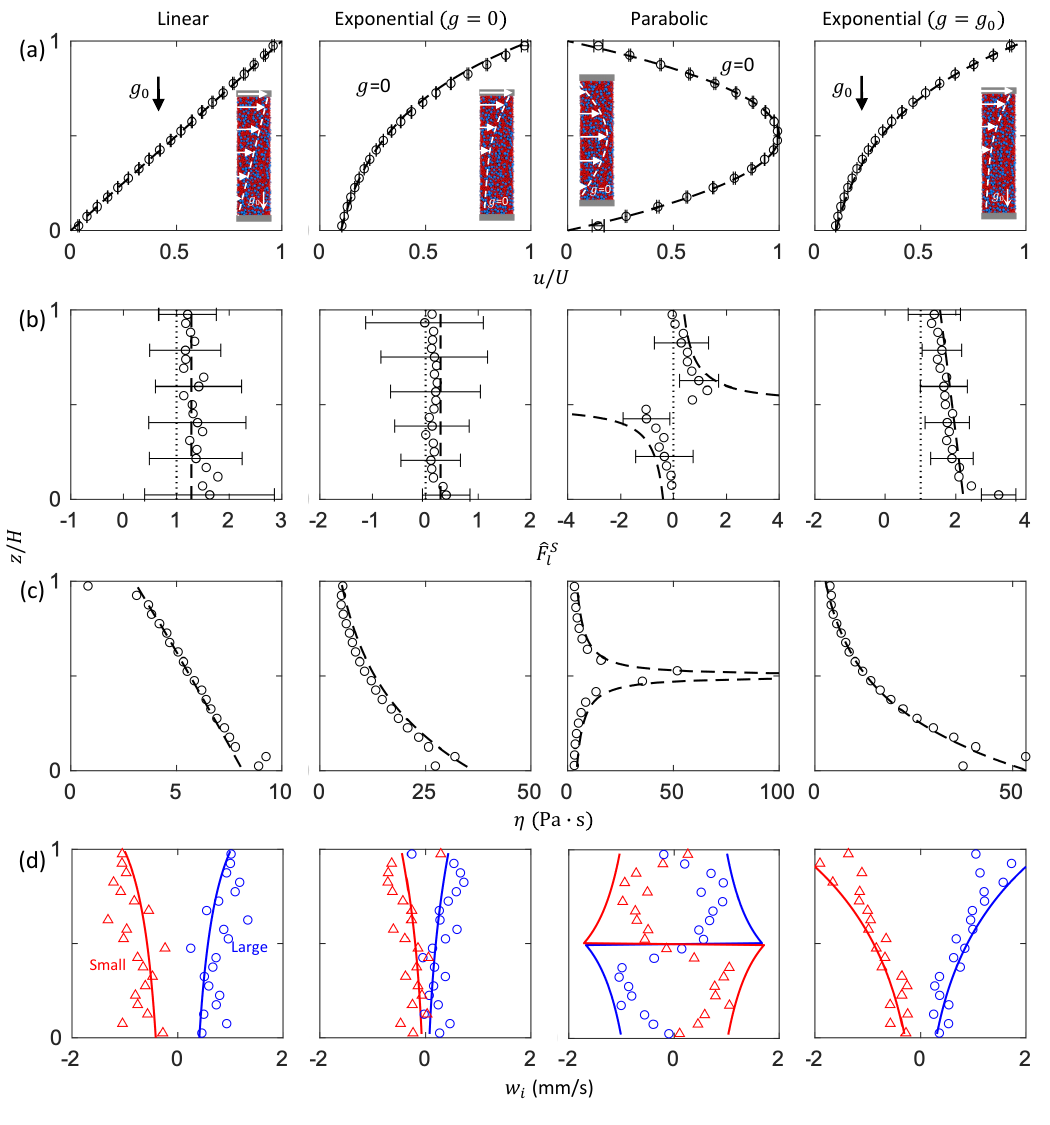}}
\caption{
\textcolor{blue}{Depth profiles (rows) of time averaged simulation results (symbols) and predictions (dashed black curves) for the four controlled shear flows (columns) in steady state at $R_d=2$.} (a) Streamwise mean velocity $u$, (b) \textcolor{blue}{normalized} segregation force on a large particle \textcolor{blue}{$\hat F^S_l=F^S_l/m_l g_0$}, (c) bulk viscosity $\eta$, and (d) segregation velocity, $w_i$ for small (red) and large (blue) particles measured from the simulation (symbols) and predicted via (\ref{eq:w}) (curves).  Dotted vertical lines in (b) indicate segregation force equal to particle weight.
In all cases, $U=20\,$m\,s$^{-1}$, $c_l=c_s=0.5$, and $H\approx 0.2\,$m.
}
    \label{fig:controlled}
\end{figure}

The simulation results and predicted values for the segregation velocity for the four controlled shear flows at $R_d=2$ are presented in figure~\ref{fig:controlled}.
Row (a) shows the prescribed and measured streamwise velocity profiles along depth, $z$. 
The close agreement between the DEM data points and the curves representing the target velocity profile demonstrates the effectiveness of the control scheme in achieving the desired velocity profiles.

The prescribed velocity functions in figure~\ref{fig:controlled}(a) are used to calculate the $z$-profiles of $\dot\gamma$ and $\partial \dot\gamma/\partial z$, while the pressure is estimated based on an idealized hydrostatic pressure, $p=P_0+\rho\phi g (H-z)$. 
These results are then used to calculate the predicted mixture segregation force, shown in row (b), \textcolor{blue}{which we normalize with the particle weight, $\hat F^S_i=F^S_i/m_i g_0$. The }dashed curves represent $\hat F^S_l$ calculated using (\ref{eq:intruder}-\ref{eq:mixture_tanh1}) based on the prescribed pressure and shear states. The symbols indicate $\hat F^S_l$ measured directly from the simulation using the extended virtual spring approach, where forces proportional to the offset of the center of the mass between the two species are applied uniformly to particles of each species to suppress segregation \citep{duan2024general}. Error bars indicate the standard deviation of DEM data averaged over the 1\,s measurement window.
The agreement between the DEM data and the model predictions for $\hat F^S_l$ across all four controlled shear flows confirms the general applicability of the segregation force model (\ref{eq:mixture_tanh1}) (see also \cite{duan2024general}). Results for $\hat F^S_s$ are similar (not shown). Note that the total segregation force equals the \textcolor{blue}{depth-wise component of the} particle weight for steady flows, specifically, $c_l \hat F^S_l + c_s \hat F^S_s=\cos\theta$, where the value on the RHS of this equation is shown as a vertical dotted line in row (b) of the figure. The large difference between measured and predicted values of $\hat F^S_l$ in the vicinity of $z/H=0.5$ for the parabolic velocity profile occurs because $\dot\gamma(H=0.5)=0$, leading to a singularity in the second term in (\ref{eq:intruder}), see \citep{duan2024general}\textcolor{blue}{. However, this does not affect the segregation velocity because of a corresponding singularity in the viscosity, as explained next}.

The \textcolor{blue}{effective granular} viscosity, $\eta$, measured as the ratio of shear stress $\tau$ to shear rate $\dot\gamma$\textcolor{blue}{, (\ref{eq:mu}),} averaged over the 1\,s window from DEM simulation data, is plotted in row (c) of figure~\ref{fig:controlled}.
The dashed curve represents $\eta$ estimated from the $\mu(I)$ rheology (\ref{eq:viscosity}) for the corresponding prescribed pressure-shear state.
The predicted and measured values of viscosity match well except near the flow boundaries (see appendix~\ref{appendixA}).
Similar to the segregation force, the viscosity exhibits a singularity at $\dot\gamma \approx 0$ for the parabolic velocity profile. 
This singularity, however, is eliminated as $\dot\gamma \to 0$ when the segregation force, along with the drag force, are incorporated into the force and momentum balance, because both forces scale as $1/\dot\gamma$.


With the knowledge of segregation force (row b), viscosity (row c), and drag coefficient ((\ref{eq:drag_coefficient}) for large particles and (\ref{eq:conservation}) for small particles), equation (\ref{eq:w}) allows the calculation of segregation velocity. Moreover, the depth-dependent segregation velocity can be directly measured for each horizontal layer by quantifying the rate of center of mass offset between the two species (\ref{eq:com}) in a 1\,s window after the flow reaches steady state.  
Figure \ref{fig:controlled}(d) shows both the model predictions and DEM measurements of the segregation velocity for the four controlled shear flows. 

For the linear velocity profile (first column), the segregation velocity decreases from top to bottom, despite a constant depth-independent segregation force. This decrease is due to the increase of pressure with depth, which increases the viscosity term in the drag force, thereby slowing the segregation.
A similar pattern is observed for the exponential velocity profile without gravity in the second column. 
Despite the constant segregation force and pressure, the shear rate decreases with depth, which increases the viscosity and, hence, the drag.
For the parabolic velocity profile (third column), the segregation force changes sign at $\dot{\gamma}=0$. Consequently, the segregation velocity also changes sign at this point, consistent with large particles segregating to regions of higher shear \citep{fan_phase_2011}.
While the model accurately predicts this sign change, the flow in the $z/H\approx 0.5$ region has negligible shear and, hence, is quasi-static. 
The continuum framework upon which the model is based is difficult to apply in this regime, resulting in significant discrepancies between the model predictions and the simulation measurements.
Finally, in the case of the exponential velocity profile with gravity (fourth column), the segregation force increases linearly with depth. 
However, the viscosity experiences a more pronounced increase, resulting in a decrease in segregation velocity with depth.

Overall, the consistent agreement between the model predictions and the simulation results across all four cases in both magnitude and sign  demonstrates the effectiveness of (\ref{eq:w}) in predicting segregation velocity based on the local pressure-shear state. 
These results further indicate that the model effectively captures the underlying physical mechanisms driving segregation and the associated segregation velocity.  The match between the measured segregation velocity and its predicted values in figure~\ref{fig:controlled}(d) is \textcolor{blue}{not perfect}. Nevertheless, given the relatively straightforward model, the simplifying assumptions used in the model, the difficulty in isolating the various forces acting on individual particles, the complication of considering mixtures (rather than an intruder), and the inherent stochastic nature of granular flows, the match between the measured and predicted segregation velocities is surprisingly good.

\subsection{Varying species concentration}

\begin{figure} \centerline{\includegraphics[width=\columnwidth]{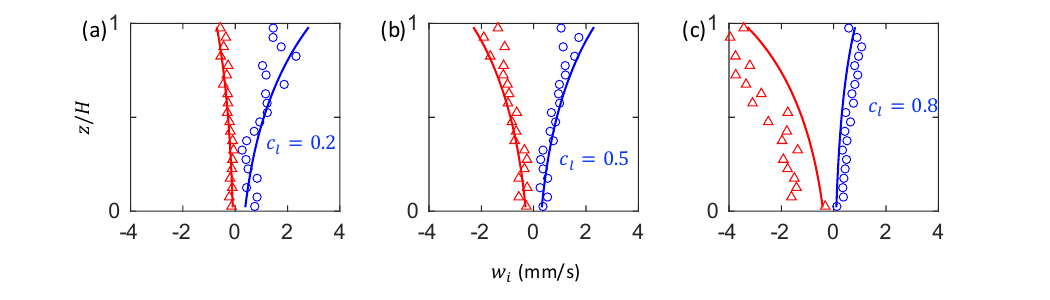}}
\caption{
\textcolor{blue}{Profiles of the segregation velocity $w_i$} for large (blue) and small (red) particles with $R_d=2$ for the exponential velocity profile with $g=g_0$ and bulk large particle concentrations of (a) $c_l=0.2$, (b) $c_l=0.5$, and (c) $c_l=0.8$, based on the prescribed velocity profiles (solid curves) compared to DEM measurements (symbols) averaged over 1\,s after the flow reaches steady state. 
}
    \label{fig:controlled_differentc}
\end{figure}

The analysis in section~\ref{controlled} focuses on uniformly mixed systems with equal volume fractions of small and large particles, $c_l=c_s=0.5$. 
However, the concentration dependence of the segregation velocity as described by (\ref{eq:w}) should  hold for any species concentration within the range $0\le c_l\le1$, where $c_s=1-c_l$.
To verify this, we consider controlled shear flows with the exponential velocity profile and $g=g_0$, because for these conditions both the segregation force and the viscosity vary with depth (see the last column of figure~\ref{fig:controlled}).
Figure~\ref{fig:controlled_differentc} shows the model predictions for $w_i$ at $R = 2$  with uniform species concentrations $c_l = 0.2$ and 0.8, along with the results for $c_l = 0.5$ (repeated from figure~\ref{fig:controlled} with a different horizontal scale). 
While a wider scatter of data points is observed with $c_l=0.2$, the predicted segregation velocity matches the measured segregation velocity reasonably well.  Furthermore, a general trend of decreasing segregation velocity with depth is evident for both particle species, similar to $c_l=0.5$. However, the segregation velocity for large particles increases for $c_l=0.2$, and the segregation velocity for small particles decreases, compared to $c_l=0.5$.  That is, for $c_l=0.2$ a large particle among mostly small particles segregates faster than a small particle among few large particles. In contrast, for $c_l=0.8$, the model under predicts the segregation velocity, possibly due to the difficulty in accurately determining $\hat F^S_{s}$ for small intruders in the sea of large particles, which is highly sensitive to $R_d$ and $c_l$ near the monodisperse limit of $c_l=1$~\citep{jing_unified_2021}.  Nevertheless, the decreased segregation velocity with depth and the increased (decreased) segregation velocity for small (large) particles compared to $c_l=0.5$ is clear.  Here, a small particle among many large particles segregates faster than a large particle among many large particles, consistent with previously observed asymmetry in the segregation velocity \citep{van_der_vaart_underlying_2015,jing_micromechanical_2017,jones2018asymmetric}. \textcolor{blue}{This asymmetry in the segregation velocity originates as an asymmetry in the segregation force (\ref{eq:mixture_tanh1}), which in turn can be traced back to segregation being intruder-like for small particles at large $c_l$ and intruder-like for large particles at small $c_l$. The species concentration range that the segregation force is intruder-like is narrower for small particles and wider for large particles (\cite{duan_segregation_2022}, figures 2 and 3), which is reflected in the hyperbolic tangent dependence of the segregation force on concentration (\ref{eq:mixture_tanh1}) and, consequently, in the form of the segregation velocity (\ref{eq:w}). It is also consistent with the kinetic sieving model of \cite{savage1988particle}, as demonstrated by \cite{jones2018asymmetric} by comparing DEM segregation velocity results to the solution of the Savage and Lun model over a range of size ratios and species concentrations (\cite{jones2018asymmetric}, figures 5 and 7).}

To this point, the segregation velocity model (\ref{eq:w}) has been validated against different flow configurations with uniform mixture concentrations throughout the flow domain.
However, the model can be extended to scenarios where particle species concentration varies with depth.
In such cases, concentration gradients induce a diffusion flux that can enhance or counteract the segregation flux, as described in section~\ref{section:diff}. 
Because differentiating between these two fluxes is impractical from a measurement standpoint, the observed segregation velocity represents the net effect of both fluxes (see (\ref{eq:w_net})). 


\begin{figure} \centerline{\includegraphics[width=\columnwidth]{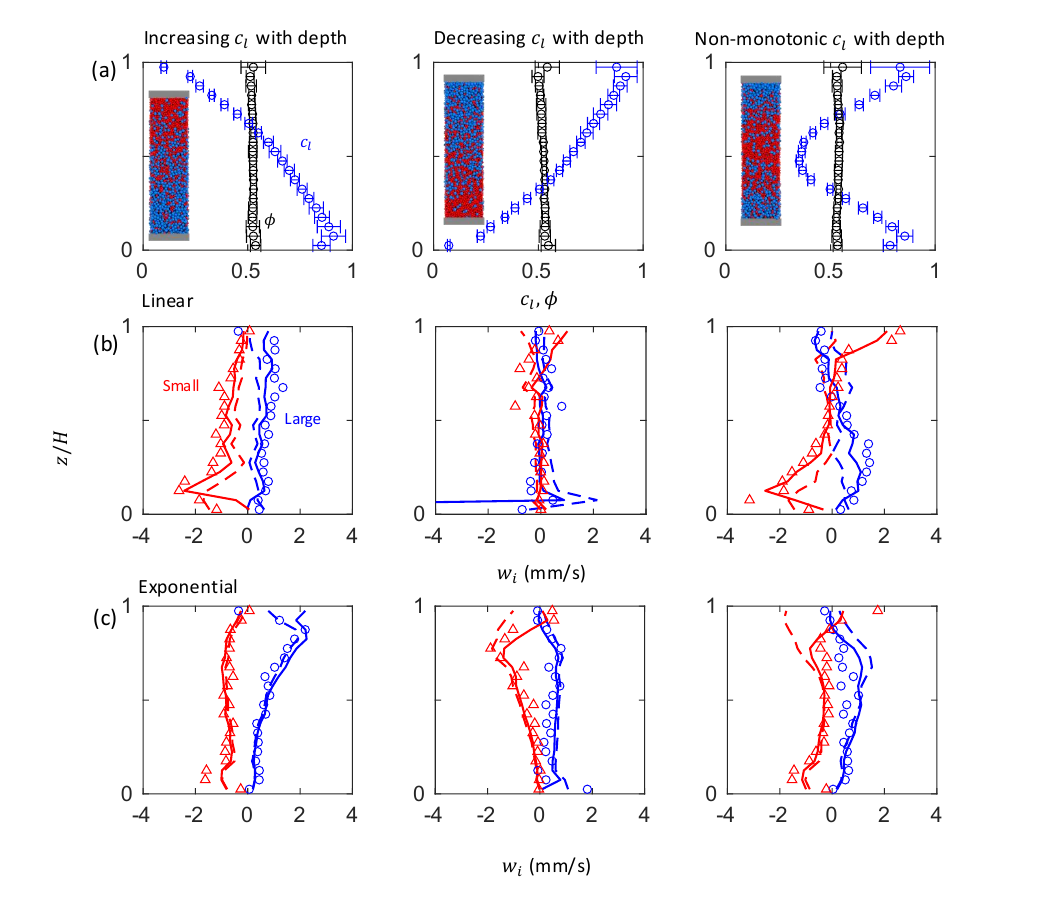}}
\caption{
Effect of (a) three different spatially varying concentration profiles (columns) on the segregation velocity \textcolor{blue}{$w_i$} for (b) linear ($u=Uz/H$) and (c) exponential ($Ue^{k(z/H-1)}$) velocity profiles with $g=g_0$ and $R_d=2$. In rows (b) and (c), dashed curves represent model predictions using equation (\ref{eq:w}) for $w_i$, solid curves represent predictions corrected by the diffusion flux, i.e.\ $w^{net}_i$ from (\ref{eq:w_net}), and symbols indicate measurements from DEM simulations.
}
    \label{fig:controlled_varyingc}
\end{figure}

To consider varying species concentrations, three distinct large-particle concentration profiles, shown in row (a) of figure~\ref{fig:controlled_varyingc},  are investigated: $c_l$ increasing  with depth,  $c_l$ decreasing  with depth, and $c_l$ decreasing in the upper half of the flow and increasing in the lower half of the flow. 
Although these concentration profiles are initialized to vary linearly with depth, variations in concentration occur as the flow is established, resulting in slightly nonlinear $c_l$ profiles at the start of the 1\,s measurement window. The particle volume fraction, $\phi$, remains approximately constant with depth in all three cases.
Row (b) shows the segregation velocity for an imposed linear velocity profile with $g=g_0$ in the $z$-direction. 
Neglecting the diffusion flux leads to an underestimation of the segregation velocity \textcolor{blue}{(dashed curves)} for both small and large particles, as both the diffusion flux and the segregation flux act in the same direction. 
Specifically, both segregation and diffusion contribute to the upward movement of large particles. 
The improved agreement between the diffusion-corrected velocity ($w_i^{net}$, solid curves) and the DEM measurements shows the importance of the diffusion effect in  inhomogeneous-concentration flows across all three initial conditions for $c_l$. Results for an exponential velocity profile with $g=g_0$ in the $z$-direction in row (c) indicate a generally good agreement between the predicted segregation velocity and the DEM results even without accounting for the diffusion flux, except near the top boundary where the correction for diffusion flux is clearly necessary.  This is because the diffusion flux scales with shear rate, which is maximal at the top boundary and decreases rapidly with increasing depth.

For all cases in figure~\ref{fig:controlled_varyingc}, the predicted segregation velocity accounting for diffusion matches the measured segregation velocity well.
It is also worth noting that the force-based segregation velocity model accurately captures the change in segregation direction in scenarios where $c_l$ reaches a minimum value at a mid-depth position for both linear  and exponential  velocity profiles. 
This not only highlights the model's ability to predict segregation velocities, even with substantial variations in the concentration gradient, but also reveals that diffusion can, under certain conditions, become the dominant mechanism, overpowering the effects of size-induced segregation.

\subsection{Natural flows}

\begin{figure} \centerline{\includegraphics[width=\columnwidth]{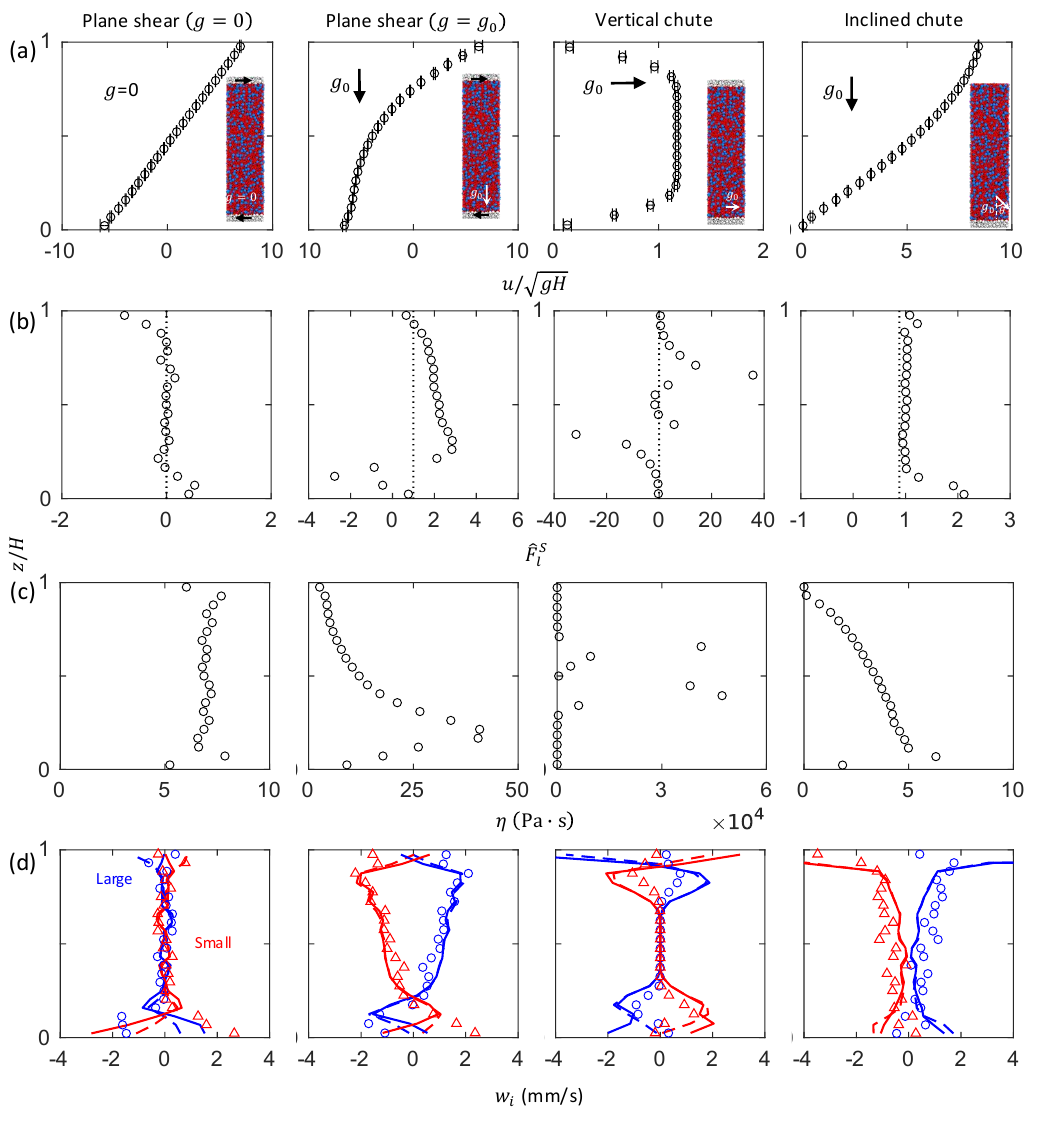}}
\caption{Depth profiles (rows) of time averaged simulation results (symbols) for the four natural shear flows (columns) in steady state at $R_d = 1.5$. (a) Streamwise mean velocity $u$, (b) \textcolor{blue}{normalized} segregation force on a large particle \textcolor{blue}{$\hat F^S_l=F^S_l/m_l g_0$}, (c) bulk viscosity $\eta$, and (d) segregation velocity \textcolor{blue}{$w_i$} for small (red) and large (blue) particles measured from the simulation (symbols) and predicted by (\ref{eq:w}) (dashed curves) and considering diffusion  (\ref{eq:w_net}) (solid curves).  Dotted vertical lines in row (b) indicate segregation force equal to particle weight. In all cases, $c_l=c_s=0.5$ and $H\approx0.2\,$m.
}
    \label{fig:natural}
\end{figure}

As demonstrated above, the segregation velocity model (\ref{eq:w}) works well for arbitrary concentration fields in a variety of flows where the velocity field is artificially prescribed.  We now examine four uncontrolled wall- or gravity-driven flows, illustrated in row (a) of figure~\ref{fig:natural}, in which the velocity field develops naturally via the boundary conditions and gravity-induced body forces.
Note that particles in each layer of the flow are constrained vertically using the restoring force approach so that they cannot segregate until they are released when the velocity profile reaches steady-state.
In wall-driven plane shear flow, which results from upper and lower walls moving in opposite directions with velocity $U=10$\,m\,s$^{-1}$, there is no gravity, so the fully developed velocity profile is linear with a nearly constant inertial number of 0.2 \citep{duan2024general}.  
In the three gravity-driven cases, the orientation of gravity relative to the flow direction is parameterized by $\theta$. For wall-driven plane shear flow, gravity acts perpendicular to the flow direction (aligned with the $z$-direction such that $\theta=0$), resulting in a nonlinear velocity profile due to the increasing pressure with depth in the flow. 
In vertical chute flow, gravity aligns with the $z$-direction ($\theta=\pi /2$), such that there is no pressure gradient in the $z$-direction, $\partial p/\partial z=0$, resulting in a blunt velocity profile with zero velocity at the two confining walls. 
Finally, for inclined chute flow, $\theta=28^\circ$, which exceeds the critical angle required for flow. The resulting velocity profile is nonlinear with a maximum velocity at the free surface. 

Following a similar methodology to the analysis of controlled-velocity flow fields, we plot dimensionless depth profiles of $u$, $\hat F^S_i$, $\eta$, and $w_i$  for the natural flows with $c_l=c_s=0.5$, as shown in figure~\ref{fig:natural}.
Here we consider $R_d=1.5$ to demonstrate results for a different size ratio than used above.  Unlike the controlled flows where $\dot\gamma$, $\partial \dot\gamma/\partial z$, $p$, and $\eta$ can be determined  from  prescribed functions (streamwise velocity, pressure) or assumed constant (concentration), these same variables for natural flow are directly measured from DEM simulations.

In the absence of an intrinsic velocity scale for vertical and inclined chute flows, the kinematic terms are nondimensionalized using the acceleration due to gravity at the earth's surface ($g_0$) and the flow depth ($H$). 
Unlike the controlled flows in the previous sections, here there are no predicted values for $\hat F^S_i$ and $\eta$ because $u$ and, hence, $\dot\gamma$, $\partial \dot\gamma/\partial z$, $p$, and $\eta$ depend on the naturally developed flow conditions, rather than being prescribed.  Nevertheless, the predicted value for the segregation velocity, $w_i$, based on measured $z$-dependent values of $\dot\gamma$, $\partial \dot\gamma/\partial z$, $p$, and $\eta$, can be compared to the measured value of $w_i$.

Consider first the wall-driven plane shear flow without gravity (column 1 of figure~\ref{fig:natural}) where the streamwise velocity decreases linearly with increasing depth, resulting in a constant shear rate. This leads to a negligible segregation force across the depth, with only minor deviations observed near the walls. Furthermore, the viscosity remains relatively constant throughout the flow domain. Consequently, the segregation velocity in row (c) of figure~\ref{fig:natural} is approximately zero over most of the depth, except close to the walls.
The minor deviation from perfect symmetry about $z/H=0.5$ near the top and bottom walls 
is likely a consequence of two factors.
First, the initial packing procedure, which involves particles falling freely under gravity in the negative $z$-direction to fill the domain, may lead to a slight initial segregation of particles.
 Second, random variations in roughness between the top and bottom boundaries could also contribute to the asymmetry.

With gravity (column 2), the wall-driven plane shear flow velocity profile is steep near the upper moving wall but flattens in the bottom half of the flow where the pressure is higher. 
As a result, the segregation force increases with depth except within the region $z/H<0.3$. In this region, a rapid decrease in segregation force is observed, accompanied by a reversal in its sign at approximately  $z/H\approx 0.2$. This is due to the interaction of particles with the moving bottom wall, leading to complex variations in the shear rate gradient for small $z/H$.
A similar trend also occurs for the viscosity.
Consequently, the segregation velocity in row~(d) of figure~\ref{fig:natural} demonstrates a nonlinear variation with depth, with a reversal in segregation direction at $z/H\approx 0.2$.
Nevertheless, the predicted segregation velocity matches the measured values well except where it is affected by the walls.

The vertical chute flow (column 3) has a plug-like velocity profile, resulting in a segregation force that is antisymmetric about $z/H=0.5$. 
Within the central region of the flow, specifically $0.2\le z/H\le 0.8$, the shear rate is negligible, leading to difficulty in accurately estimating the local viscosity, which is typical of the quasi-static flow regime.
Despite this problem, the model accurately predicts the measured segregation velocity to be approximately zero within this quasi-static central region, as well as the reversal of segregation direction above and below this region. 

The curvature of the velocity profile for the inclined chute with $\theta=28^\circ$ (column 4) is opposite that of the wall-driven flow with gravity
(column 2). The combined effects of shear rate and shear rate gradient result in an almost uniform segregation force except near the static bottom boundary.
However, the increase in viscosity with depth leads to a gradual decrease in the segregation velocity, except near the bottom wall, as shown in row (c) of figure~\ref{fig:natural}. 
Additionally, the model over-predicts the segregation velocity near the free surface ($z/H \ge 0.9$).
This discrepancy is attributed to the inherent difficulty in accurately resolving the free surface using bin-averaging, compounded by the dilute flow regime ($I>5$) that prevails near the free surface. Such conditions deviate from the dense flow regime for which the model was developed, resulting in the observed over-prediction of $w_i$.
Nevertheless, through most of the depth of the chute flow, the predicted segregation velocity is consistent with the measured values.

\begin{figure} \centerline{\includegraphics[width=\columnwidth]{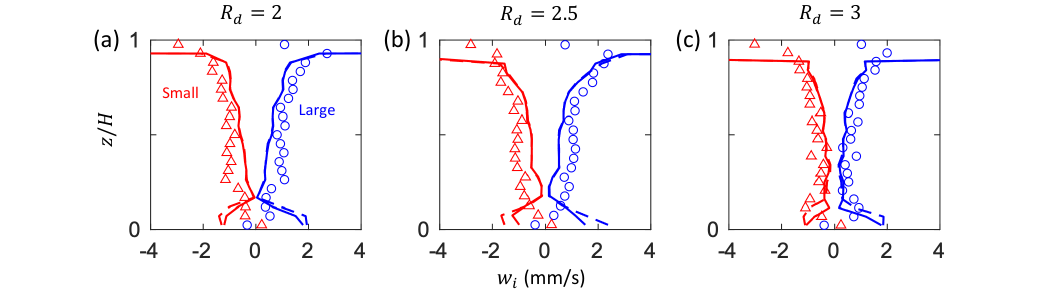}}
\caption{
Segregation velocity \textcolor{blue}{$w_i$} for small (red) and large
(blue) particles measured from the simulation (symbols) and predicted via (\ref{eq:w}) (dashed curve) and after
considering diffusion via (\ref{eq:w_net}) (solid curve) for chute flow inclined at $28^\circ$ with different size ratios. In all cases, 
$c_l = c_s = 0.5$ and $H\approx 0.2\,$m.
}
    \label{fig:chute}
\end{figure}

To further confirm the validity of our approach for calculating the segregation velocity, model predictions are compared to chute flow simulations for size ratios other than $R_d=1.5$. 
Figure~\ref{fig:chute} shows reasonably close agreement between the model predictions and DEM measurements for $R_d =2$, $2.5$, and $3$, although the model under-predicts the segregation velocity.
The overall segregation velocity is largest for $R_d=2$ and $2.5$, but is smaller for $R_d=1.5$ and $3$, consistent with previous findings \citep{alonso1991optimum,felix2004evidence,thornton2012modeling,jones2018asymmetric}. 
Similar to the results for $R=1.5$ in figure~\ref{fig:natural}, the discrepancy for $z/H\ge 0.9$ is a result of dilute flow near the free surface. 
Nevertheless, the agreement observed between the model predictions and the simulation results, not only for various natural flow configurations but also across different size ratios, shows the broad applicability of (\ref{eq:w}) to a wide range of size-bidisperse granular flows at inertial numbers typical of dense flows.

\section{Comparison with a previous segregation-velocity model}
\label{sec:heap_comparison}

The broad applicability of (\ref{eq:w}) can be further elucidated by demonstrating its compatibility with the well-established linear segregation velocity model, initially developed for free surface heap flows \citep{fan2014modelling}, and successfully applied to a wide range of free surface flows including chute flow, rotating tumbler flow, steady and intermittent heap flow, three-dimensional heap flow, multidisperse species, polydisperse species, and density-disperse species \citep{umbanhowar2019modeling}.  The model \textcolor{blue}{is expressed as ~\citep{fan2014modelling}}
\begin{equation}
    \textcolor{blue}{w_i=S d_s \dot\gamma (1-c_i).}
    \label{eq:linear}
\end{equation}
It postulates a direct proportionality between the segregation velocity \textcolor{blue}{$w_i$} and the product of the concentration complement, $1 - c_i$, the shear rate, $\dot\gamma$, \textcolor{blue}{and the particle size, $d_s$ (or $d_l$), and it is motivated by the ``kinetic sieving'' mechanism and statistical mechanics model for dense granular flows of bidisperse mixtures of spherical particles by \cite{savage1988particle}, as noted previously~\citep{gray2005theory, jones2018asymmetric}. }The segregation parameter $S$ multiplied by the small particle diameter $d_s$ quantifies the characteristic segregation length scale. Larger values of $S$ indicate a stronger tendency for segregation. As determined empirically, this length scale depends on various factors, including particle properties (such as size, shape, density), flow conditions (such as flow depth), and the specific segregation mechanism at play. The effectiveness of the linear segregation velocity model stems from its ability to encapsulate the complex interplay of segregation factors in a single length scale, while still demonstrating good agreement with simulations and experiments of free surface flows on heaps or in rotating tumblers. For free surface flows without strongly segregated regions, the empirically determined segregation length scale $S$ is well approximated by~\citep{schlick2015modeling}
\begin{equation}
    S=0.26 \ln R_d.
    \label{eq:fan}
\end{equation}
 However, the assumption of a linear relationship between segregation velocity and shear rate in (\ref{eq:linear}), while simple to implement, neglects the influence of pressure.  Consequently, this linear segregation model is limited to free-surface flows \textcolor{blue}{where the effects of lithostatic pressure on its predictions are negligible}. 

     \label{eq:fry}

\begin{figure} \centerline{\includegraphics[width=\columnwidth]{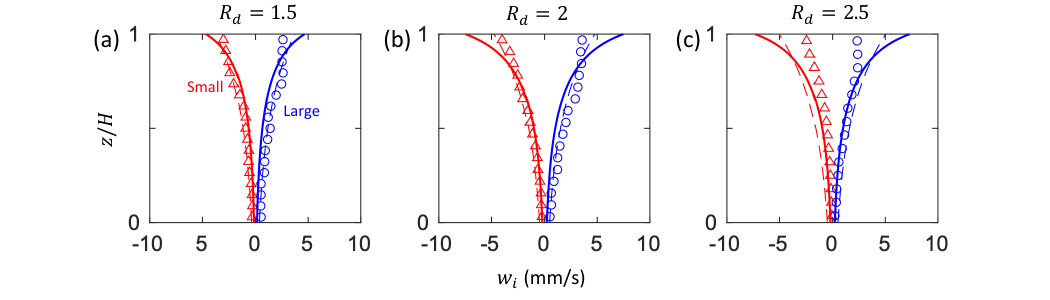}}
\caption{
\textcolor{blue}{Profiles of the segregation velocity $w_i$} for large (blue) and small (red) particles at the feed zone exit of 
quasi-2-D heap flows for three different size ratios: (a) $R_d=1.5$, (b) $R_d=2$, and (c) $R_d=2.5$.  Curves represent predictions from (\ref{eq:linear}) (dashed) and (\ref{eq:w}) (solid). Symbols represent DEM measurements averaged over 1\,s after the flow becomes steady. 
}
    \label{fig:comparison}
\end{figure}

To demonstrate that the force-based segregation velocity model described in this paper is consistent with this previous simpler approach for free-surface flows, model predictions from equations (\ref{eq:linear}) and (\ref{eq:w}) are compared to data from quasi-two-dimensional heap flow simulations, which are detailed elsewhere 
\citep{duan_modelling_2021}.
The simulations are performed with a two dimensional flow rate of 20\,cm$^2$\,s$^{-1}$ and a heap length of 52\,cm, resulting in a measured flowing layer thickness ($\delta$) of approximately 1.5\,cm. The large particle diameter is 3\,mm, while the small particle diameter is varied between 1.2\,mm and 2\,mm, to provide different size ratios. 
The feed concentration is $c_l=c_s=0.5$,  which is used as the input concentration for the model.
To mitigate the impact of segregation on the concentration profiles, measurements of the segregation velocity are taken immediately downstream of the feed zone, where the concentration profile remains largely unaffected by segregation, and are averaged over a 1\,s window. 

Previous studies of quasi-2-D heap flow kinematics \citep{fan_kinematics_2013} show that the streamwise velocity just downstream of the feed zone exit is well approximated by
\begin{equation}
    u(z)=\frac{kq}{\delta(1-e^{-k})}e^{kz/\delta},
\end{equation}
where $k=2.3$.
Hence, we use this velocity profile to determine $\dot\gamma$ and $\partial \dot\gamma/\partial z$, and we assume the pressure at the free surface to be the pressure of a single layer of particles, $p_0=\rho\phi g \bar d$.
The pressure profile is $p(z)=\rho\phi g (\bar d +\delta-z) \cos\theta$, and the feed concentrations of the two particle species are $c_l=c_s=0.5$. 

Predictions of the linear segregation-velocity model (\ref{eq:linear}) and the force-based model (\ref{eq:w}) are generally consistent with each other for all three size ratios, as shown in figure~\ref{fig:comparison}. Both methods also match measurements of the local segregation velocity from the simulations, except near the free surface, particularly for $R_d=2$ and $2.5$.
Note that in all cases, the small and large particles are assumed to be perfectly mixed with a uniform concentration of $c_l=c_s=0.5$ at this location in the flow, which implies equal but opposite segregation velocities for the two particle species. 
This is indeed the case for the measured segregation velocity for $R_d=1.5$ and 2, where the segregation velocities for both species are nearly symmetric about zero.
However, for $R_d=2.5$, the asymmetry in the measured segregation velocities of the two particle species suggests that some segregation has occurred in the feed zone.
In this case, while both (\ref{eq:linear})  and (\ref{eq:w}) accurately predict the segregation velocity of the large particles, they tend to overestimate the segregation velocity of the small particles. 

In all cases in figure~\ref{fig:comparison}, the force-based model over-predicts the segregation velocities in the near-surface region ($z/H>0.9$). 
Similar to the chute flow results in figure~\ref{fig:chute},
this discrepancy is likely attributed to the dilute flow with diminishing pressure near the free surface, as well as the additional challenges associated with delineating the free surface when employing a cut-off solid fraction as the defining criterion \citep{duan_modelling_2021}. 
Nevertheless, the close agreement between the force-based model (\ref{eq:w}) and the linear segregation velocity model (\ref{eq:linear}), along with their collective agreement with the measured segregation velocity for $z/H<0.9$, provides compelling evidence that the two models are consistent with one another and that the linear segregation model provides a reasonable simplification for heap flow and other free surface flow scenarios.

\textcolor{blue}{
Finally, we note that other segregation velocity models exist, but they are limited in several ways.  For one, the effect of shear rate gradients on segregation is not included in many models and scalings~\citep{marks2012grainsize,tunuguntla2014mixture,fry2018effect,chassagne_discrete_2020,trewhela_experimental_2021}, making them inapplicable to most of the flows in this study. Other approaches are more general but lack details (such as particle size and concentration dependence) that would allow them to be applied to the specific flow situations considered here~\citep{hill_segregation_2014,gray2015particle,bancroft_drag_2021,liu2023coupled,singh2024continuum}.
Hence, even though some of these segregation velocity models and scalings have demonstrated success for specific flow conditions and geometries, it would be challenging to meaningfully evaluate their relative performance across the broad range of flow conditions considered here.
}

\section{Conclusions}
The results in this paper complete our decade-long quest to predict particle segregation in dense granular flows.  This effort began using a simple approach to predict size segregation for bidisperse mixtures of dense flowing particles using an advection-diffusion equation (\ref{transport1}) that includes a term related to segregation \citep{fan2014modelling,umbanhowar2019modeling}. This approach was first proposed four decades ago \citep{bridgwater1985particle} and has been built upon by many other researchers \citep{gray2005theory,marks2012grainsize,tunuguntla2014mixture,hill_segregation_2014}.  However, the key to its practical implementation was a simple expression for the segregation velocity (\ref{eq:linear}), which was motivated by the much more complicated “kinetic sieving” model of \cite{savage1988particle}.  Using (\ref{eq:linear}) for the segregation velocity allows the application of the advection-diffusion-segregation model across a wide range of surface flow geometries including heap flow, tumbler flow, chute flow, and hopper flow \citep{fan2014modelling,schlick2015granular,deng2020modeling}.  It can be applied not only to bidisperse granular materials, but also multidisperse and polydisperse particle mixtures \citep{deng2018continuum,gao2020broad}, density-disperse granular materials, where $S$ depends on the particle density ratio instead of the size ratio, combined size- and density-disperse granular materials \citep{duan_modelling_2021}, and even non-spherical particles \citep{jones2020remarkable, jones2021nonspherical}.  And, while it is best suited for segregation in free-surface flows, adjustments can be made such that it can be applied in situations with significant overburden pressures \citep{fry2018effect}.

While successful in many ways, the advection-diffusion-segregation approach using (\ref{eq:linear}) for the segregation velocity has been limited by the empirical basis (\ref{eq:fan}) for the segregation length scale, $S d_s$, which is a function of the size or density ratio, as determined from DEM simulations, although it is possible to estimate $S$ by matching experiments to predictions using the advection-diffusion-segregation equation \citep{fry2020measuring1,fry2020measuring2}.  Hence, we set out to determine the segregation velocity based on particle-level forces.  This requires several ingredients.  Foremost is an understanding of the forces acting on an individual particle in a granular flow.  While seemingly straightforward, it was not until the development of a virtual spring approach for DEM simulations \citep{guillard_scaling_2016} that it became possible to extract meaningful measurements of forces on an intruder particle in a dense granular flow.  This allowed a better understanding of the segregation force acting on the intruder particle, which is a combination of buoyancy-like effects and gradients in the shear rate, expressed compactly in (\ref{eq:intruder}).  A related approach provided the means to measure the drag force on a particle pulled through a sheared granular medium and determine that the drag on an intruder particle is Stokes-like over a range of Reynolds numbers extending over several orders of magnitude \citep{jing_drag_2022,he2025lift}, expressed in (\ref{eq:drag}), with the drag coefficient dependent on the intruder size ratio and density ratio as well as the inertial number.  An additional ingredient in the drag force formulation is the ability to determine a granular viscosity via the $\mu(I)$ rheology (\ref{eq:mu_eff}) \citep{midi2004dense}. The segregation force, drag force, and particle weight must balance for an individual intruder particle under the assumption of negligible acceleration, which allows the determination of its segregation velocity, much like how the terminal velocity of a sphere falling in a viscous fluid can be determined from the buoyancy force, drag force, and weight.  However, as in a fluid suspension where nearby particles and the local flow kinematics alter the sedimentation velocity, other nearby intruder particles and the local flow conditions alter the segregation velocity for a particle in a flowing granular mixture.  Using a variant of the virtual spring approach \citep{duan_segregation_2022,duan2024general}, the dependence of these forces on species concentration can be determined as given by (\ref{eq:mixture_tanh1}).  The final ingredient corrects the segregation velocity in a mixture to account for diffusion fluxes originating from concentration gradients, given by (\ref{eq:w_net}).  

As we show in this paper, properly combining all of these ingredients results in the ability to predict the segregation velocity using forces rather than calculating it from the simple relation of (\ref{eq:linear}), which depends on an empirical relation for the segregation length scale, $S d_s$, and is limited to free surface flows.  This new approach for determining the segregation velocity matches the measured segregation velocity well for the full variety of flow and mixture conditions examined here. 


Although determining the segregation velocity using particle-level force models allows the application of the advection-diffusion-segregation equation across a wider range of conditions, there is more work to be done. \textcolor{red}{For instance, predictions of segregation from the force-based approach could be compared to previously proposed segregation kinematics models for specific flow and material combinations (e.g.,  \cite{fry2018effect,rousseau_bridging_2021,trewhela_experimental_2021,jing_drag_2022,singh2024continuum})}. Extensions to polydisperse particle size distributions, combined size and density segregation, more complex flow geometries, and non-spherical particles are needed.  Of particular interest is increasing the size ratio range that can be accurately modelled.  The mixture segregation force (\ref{eq:mixture_tanh1}) has only been shown to be valid for $R_d<3$ and the coefficient of drag (\ref{eq:drag_coefficient}) for $R_d<5$. 
The challenge is that for large $R_d$ the nature of particle interactions changes.  In fact, for $R_d \geq 6.464$ small particles fall freely through the interstices between large particles even if the particles are not undergoing shear, a process called free sifting.  This effect begins to influence segregation for $R_d>3$ \citep{gao2023percolation,gao2024shearpercolation}, well before unhindered free sifting occurs at $R_d = 6.464$. Since many practical granular flows in industry and geophysics consist of mixtures with very broad size distributions, more work is clearly warranted to better understand and predict the segregation of ``fine'' particles with $R_d \geq 4$.  Other physical effects can also play important roles in segregation.  Of particular interest in the chemical and pharmaceutical industries is interparticle cohesion, which can result from moisture, surface roughness, electrostatic charging, stickiness, and other causes.  In some cases, cohesion reduces segregation, while in other cases fine particles agglomerate into large clusters which then segregate.  A very different issue is the coupling between the granular flow field and granular segregation, which we ignore here by simply assuming that the flow is relatively unaffected by segregation.  Progress is being made in this area \textcolor{blue}{ \citep{barker2021coupling,sahu_particle_2023,liu2023coupled,edwards2023particle,maguire2024particle,singh2024continuum},} but more work can be done.  In short, while this paper arguably closes the loop on segregation prediction for noncohesive particles with $R\lesssim 3$ by providing a particle-level force-based approach for predicting the segregation velocity, much work remains to be done for larger size ratios, cohesive particles, coupling flow to segregation, and many other practical aspects of granular segregation.

\section*{Acknowledgments} 

We acknowledge the use of retrieval-augmented generative AI for language editing purposes during the drafting stage; the associated tool and prompts can be found at \href{https://github.com/19revey/LLM_paper_writing.git}{GitHub Repository}.
This material is based upon work supported by the National Science Foundation under Grant No. CBET-1929265. L.J.\ gratefully acknowledges financial support provided by the National Natural Science Foundation of China (12472412) and the Open Research Fund Program of State Key Laboratory of Hydroscience and Engineering (sklhse-2023-B-07).

\section*{Declaration of interests} The authors report no conflict of interest.

\appendix

\section{Confirmation of $\mu(I)$ parameter values}\label{appendixA}

\begin{figure}\centerline{\includegraphics[width=1.3\columnwidth]{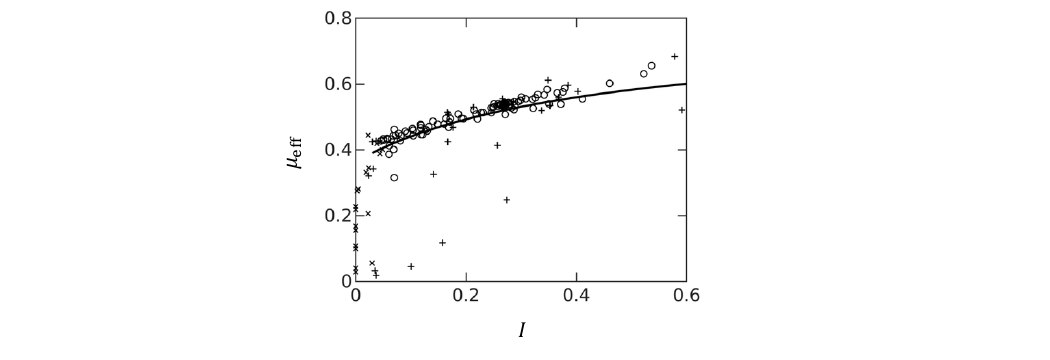}}
\caption{Effective friction coefficient $\mu_\mathrm{eff}$ vs.\ local inertial number $I$ for \textcolor{blue}{the eight controlled and natural flows} included in this study. 
The data points represent DEM simulation measurements of the ratio of shear stress to shear rate, defined as $\mu_\mathrm{eff}$.
Circles correspond to flows shown in figures~\ref{fig:controlled} and \ref{fig:natural}.
Outliers represent flows near the boundaries ($+$), where $z/H<0.1$ or $z/H>0.9$, and those in the quasi-static regime ($\times$) with $I<0.03$.
The solid curve is the prediction
of equation~(\ref{eq:mu_eff}) with $\mu_s = 0.364$, $\mu_2 = 0.772$, and $I_c = 0.434$ for data from a previous study of chute flow \citep{tripathi_rheology_2011}.
}
\label{fig:rheology}
\end{figure}

Accurate prediction of the segregation velocity from the flow field necessitates, in addition to segregation force and drag models, a value for the effective granular viscosity, $\eta$, in the drag model \ref{eq:drag}. This can be estimated from the $\mu(I)$ rheology model \textcolor{red}{for dense flows} via (\ref{eq:mu}) and (\ref{eq:mu_eff}).
The accuracy of the $\mu(I)$ rheology for the flows considered here is confirmed in figure~\ref{fig:rheology}, which shows the relationship between the local effective friction coefficient ($\mu_{\text{eff}}$) and the local inertial number ($I$) for the eight controlled and natural flows included in this study. 
Nearly all of the data falls on the master curve predicted by the $\mu(I)$ model with parameters derived from a separate study using simulations with different particle properties (density-bidisperse mixture) and flow geometry (periodic chute) \citep{tripathi2013density}. 
Notably, outlier data points correspond to locations near the solid wall boundaries or within quasi-static regions of the flow, which are expected to deviate from the $\mu(I)$ rheology. \textcolor{red}{Corrections to the $\mu(I)$ rheology model have been proposed for quasi-static flow (as $I\rightarrow 0$) and for $I \gtrsim 0.3$~\citep{barker2017well, heyman_MUI_2017, barker2021coupling,lloyd2025}} \textcolor{orange}{that involve} \textcolor{red}{fitting several parameters} \textcolor{orange}{to account for deviations from} \textcolor{red}{the standard $\mu(I)$ rheology} \textcolor{orange}{in these regions. However, these} \textcolor{red}{deviations from the $\mu(I)$ rheology model only occur near the walls and in quasi-static regions where other effects} \textcolor{orange}{also influence segregation and are not captured by our framework, so} we rely on the standard $\mu(I)$ rheology to estimate the viscosity used in the expression for the drag force, (\ref{eq:drag_approx}) and, subsequently, the segregation velocity. Finally, the viscosities estimated from (\ref{eq:mu}) compare well to the values measured directly from the simulation, as shown in figure~\ref{fig:controlled}(c).



\bibliographystyle{jfm}
\bibliography{jfm-instructions}

\end{document}